\documentstyle[12pt]{article}
\makeatletter
\@addtoreset{equation}{section}
\makeatother

\def\lb#1{\label{#1}}
\def\l#1{\lb{#1}}
\def\r#1{(\ref{#1})}
\def\c#1{\cite{#1}}
\def\i#1{\bibitem{#1}}
\def\beq{\begin{equation}}
\def\eeq{\end{equation}}
\def\bez{\begin{displaymath}}
\def\eez{\end{displaymath}}
\def\beb#1\l#2\eeb{\begin{equation}
\begin{array}{c} #1 \qquad
\end{array} \label#2  \end{equation}}
\def\bey#1\eey{\begin{displaymath}
\begin{array}{c} #1  \end{array}  \end{displaymath}}

\begin{document}

\begin{titlepage}
\begin{flushright}
November, 28, 2001\\
hep-th/0111265
\end{flushright}

\begin{centering}
\vfill
{\bf 
SEMICLASSICAL MECHANICS OF CONSTRAINED SYSTEMS
}
\vspace{1cm}

O. Yu. Shvedov \footnote{shvedov@qs.phys.msu.su} \\
\vspace{0.3cm}
{\small {\em Sub-Dept. of Quantum Statistics and Field Theory,}}\\
{\small{\em Department of Physics, Moscow State University }}\\
{\small{\em Vorobievy gory, Moscow 119899, Russia}}

\vspace{0.7cm}

\end{centering}

{\bf Abstract}

Semiclassical mechanics  of  systems  with  first-class constraints is
developed. Starting from the quantum  theory,  one  investigates  such
objects as  semiclassical states and observables,  semiclassical inner
product, semiclassical  gauge  transformations   and   evolution.   In
ordinary quantum   mechanics,   there   are  a  lot  of  semiclassical
substitutions to the Schrodinger equation (not only  the  WKB-ansatz).
All of  them  can  be  viewed as "composed semiclassical states" being
infinite superpositions of wave packets with minimal uncertainties  of
coordinates and   momenta   ("elementary  semiclassical  states").  An
elementary semiclassical  state  is  specified  by  a  set  (X,f)   of
classical variables   X  (phase,  coordinates,  momenta)  and  quantum
function f ("shape of the  wave  packet"  or  "quantum  state  in  the
background X").  A  notion of an elemantary semiclassical state can be
generalized to the constrained systems,  provided that  one  uses  the
refined algebraic  quantization  approach based on modifying the inner
product rather than on imposing the constrained conditions on physical
states. The  inner  product  of  physical  states is evaluated.  It is
obtained that classical part  of  X  the  semiclassical  state  should
belong to the constrained surface;  otherwise, the semiclassical state
(X,f) will have zero norm for all f.  Even under classical  constraint
conditions,  the semiclassical inner product is degenerate.  It can be
obtained from the refined algebraic  quantization  prescription  by  a
linearization  procedure.  One  should  factorize  then  the  space of
semiclassical  states.   Semiclassical   gauge   transformations   and
evolution of semiclassical states are studied. The correspondence with
semiclassical Dirac approach is discussed.

\vspace{0.7cm}

\vfill \vfill
\noindent

\end{titlepage}
\newpage
\sloppy

\section{Introduction}

Constrained systems  are widely investigated in modern physics.  Gauge
field theories,  quantum  gravity   and   supergravity,   string   and
superstring theories  are  examples of systems with constraints.  Only
few models are exactly solvable;  realistic physical theories requires
approximate methods.  A perturbation theory is one of such techniques,
it is usually applied to  constrained  field  systems  such  as  gauge
theories.

Another important technique of quantum mechanics and field theory is a
semiclassical approximation. Soliton quantization \c{soliton,J}, quantum
field theory in a strong external background classical field \c{GMM} or
in curved    space-time    \c{BD},    the    one-loop    approximation
\c{oneloop,Baacke}, the   time-dependent   Hartree-Fock  approximation
\c{oneloop,Hartree} and the Gaussian approximation  \c{Gauss}  may  be
viewed as examples of application of semiclassical conceptions.

There are  different  ways  of  semiclassical investigation of quantum
systems. Some of them are based on studying the asymptotic behavior of
physical quantities    as   the   parameter   of   the   semiclassical
approximation ("Planck constant") tends to zero.

The simplest Ehrenfest semiclassical method is based on  writing  down
the equations  for  average values of semiclassical observables.  Then
one makes an assumption that quantum averages tend  to  classical  and
finds that  the  quantum wave packet can move only along the classical
trajectory. Although the formal classical equations are obtained,  the
problem of  semiclassical evolution of shape of the wave packet cannot
be resolved within the Ehrenfest approach.

Many physical  quantities  can  be  expressed   via   the   functional
integrals. Calculating such integrals via the stationary-point (or the
saddle-point) technique,  one  evaluates  these  quantities   in   the
semiclassical approximation.

Another group   of   semiclassical   approaches  is  based  on  direct
substitution of  hypothetical  approximate   wave   functions   to   the
Schrodinger equation.  An  important  advantage  of such approaches is
that the accuracy of the semiclassical approximation can be  estimated
mathematically \c{Maslov1,Maslov2},  at  least  for finite-dimensional
quantum mechanical systems.  Moreover,  one can justify the  Ehrenfest
conjecture that   semiclassical   wave   packets  are  transformed  to
semiclassical under evolution.  The behavior of the shape of the  wave
packet can  be  also investigated \c{Maslov2}.  It is also possible to
say what are semiclassical states and observables.

One can try to apply semiclassical methods to several formulations  of
quantum theory  of systems with constraints.  One can use the original
Dirac approach \c{D} and consider  states  to  satisfy  not  only  the
evolution equation but also additional constraint conditions. The most
difficult problem of this quantization is construction  of  the  inner
product. One     can     impose     additional     gauge    conditions
\c{Henneaux,Faddeev} but this approach is gauge-dependent,  especially
for the case of Gribov copies problem \c{Gribov,Shabanov}.

The BRST-BFV  approach  \c{BRST,BFV}  based  on extension of the phase
space allows us to overcome difficulties of  the  Dirac  approach  and
construct a  manifestly  covariant  formulation  of  nonabelian  gauge
theories. However,  the inner product is indefinite,  so that space of
physical states should be specified by imposing the BRST-BFV condition
on physical states and factorization of state space.

An alternative way to  develop  the  quantum  theory  is  to  use  the
conception of  refined  algebraic  quantization  \c{Marolf,GM,Marolf3}
(related ideas were used in the projection operator approach \c{proj})
and modify  the  inner  product  instead  of  imposing  the constraint
conditions on states.  This gives a prescription for the inner product
in the  Dirac  approach  without  introducing indefinite inner product
spaces. The refined algebraic quantization approach seems  to  be  the
most suitable  for developing the semiclassical technique.  It is used
in the present paper.

The purposes of this paper are:

(i) to clarify a notion of  a  {\it  semiclassical}  state  (space  of
quantum states  and classical phase space are well-known notions;  the
correspondence between them is not evident);

(ii) to  show  that  there  are  equivalent  semiclassical  states,  to
investigate an  analog  of  a  notion  of  a  gauge transformation for
semiclassical mechanics;

(iii) to   investigate   semiclassical    observables    applied    to
semiclassical states;

(iv) to  study  semiclassical  transformations  ("time  evolution") of
semiclassical states;

(v) to  investigate  a  role  of  the   superposition   principle   in
semiclassical mechanics;

(vi) to  study  the  correspondence  between  the developed semiclassical
approach, Ehrenfest  and WKB-method.

This paper   is   organized  as  follows.  Section  2  is  devoted  to
investigation of a  notion  of  a  semiclassical  state.  In  ordinary
quantum mechanics, there are a lot of semiclassical substitutions that
approximately satisfy the Schrodinger equation  in  the  semiclassical
approximation. It   is   discussed   in  appendix  B  that  all  these
substitutions (including WKB-type wave functions) can be presented  as
superpositions of  wave-packet  wave  functions which can be viewed as
"elementary" semiclassical states specified  by  a  set  of  classical
variables (coordinates,  momenta,  phase) and quantum function being a
shape of the wave packet.  Constrained systems  can  be  quantized  in
different ways. It is the refined algebraic quantization approach that
allows us to introduce  elementary  semiclassical  states.  The  inner
product of   them  is  evaluated  in  section  2.  It  appears  to  be
degenarate, so  that  it  is  necessary  to  factorize  the  space  of
semiclassical states:  two  states  are  called  equivalent  if  their
dfference is of zero norm.  One can also perform gauge transformations
of semiclassical states.  They are discussed in section 3.  Sections 4
and 5 deal with semiclassical observables and evolution.  In section 6
composed semiclassical states being superpositions of wave packets are
considered. The relationship between  refined  algebraic  quantization
and Dirac approach is also discussed in section 6.  Section 7 contains
concluding remarks.

\section{Semiclassical states}

The purpose of this section is to  specify  semiclassical  states  for
constrained systems.  First  of all,  let us specify the dependence of
constraints on the small parameter of semiclassical expansion.

\subsection{Small parameter}

{\bf 1.} It is known from quantum mechanics that semiclassical methods
can be  applied  for  such  equations  that  the  coefficients  of the
derivative operators are small,  of the  order $O(h)$  as  the  small
parameter $h$ of the semiclassical expansion tends to zero,  $h\to 0$,
while the coefficients of the  multiplication  operators  are  of  the
order $O(1)$, i.e.
\beq
ih\frac{\partial \psi}{\partial t} = H(X, -ih \frac{\partial}{\partial
X}) \psi, \qquad X \in {\bf R}^n
\l{2.1}
\eeq
For quantum-field-type   equations,   it   is  convenient  to  rescale
(cf.\c{J}) the argument $X$,
\beq
X = \sqrt{h}q,
\l{2.1a}
\eeq
so that the Schrodinger equation takes the following form
\bez
i \frac{\partial  \psi}{\partial t} = \frac{1}{h} H (\sqrt{h} \hat{q},
\sqrt{h} \hat{p}) \psi
\eez
with
\bez
\hat{p} =  -i  \frac{\partial}{\partial  q},  \hat{q}  =   q,   \qquad
[\hat{p},\hat{q}]=-i.
\eez
Thus, the   semiclassical  conception  can  be  applied  even  if  the
commutator between  coordinates  and  momenta  is   not   small,   but
semiclassical observables  depend  on  the  small  parameter $h$ in an
unusual way. It is the quantum operator
\beq
\hat{H} = \frac{1}{h} H(\sqrt{h} \hat{q}, \sqrt{h} \hat{p})
\l{2.2}
\eeq
that corresponds  to  the  classical  observable $H(Q,P)$.  One should
specify the operator ordering in  eq.\r{2.2}.  For  the  definiteness,
choose the Weyl quantization of coordinate and momentum operators (see
Appendix A).  An advantage of Weyl quantization is that real classical
observables correspond to Hermitian operators.

{\bf 2.}   Consider   the  constrained  system  with  the  first-class
classical constraints:
\bez
\Lambda_a(Q,P), \qquad a=\overline{1,M}, \qquad P,Q \in {\bf R}^n.
\eez
The main requirement is that the Poisson bracket
\bez
\{ \Lambda_a, \Lambda_b \} = \frac{\partial \Lambda_a}{\partial P}
\frac{\partial \Lambda_b}{\partial Q} -
\frac{\partial \Lambda_a}{\partial Q}
\frac{\partial \Lambda_b}{\partial P}
\eez
vanishes on  the  constraint  surface $\Lambda_a(Q,P)=0$.  The quantum
constraints should depend on the  small  parameter  $h$  according  to
formula \r{2.2}.  However,  the  "quantum"  corrections  can  be  also
nontrivial, so that in general one can expect the following dependence
of quantum constraints on the small parameter:
\beq
\hat{\Lambda}_a =      \frac{1}{h}
\Lambda_a(\sqrt{h}     \hat{q}, \sqrt{h}\hat{p}) +
\Lambda_a^{(1)}(\sqrt{h}     \hat{q}, \sqrt{h}\hat{p}) + ...
\l{2.2x}
\eeq
The simplest  case  is  abelian,  when  the  quantum  constraints  are
Hermitian and commute each other not only on the constraint surface,
\bez
[\hat{\Lambda}_a, \hat{\Lambda}_b] = 0.
\eez
Consider the quantum  constrained  system  in  the  refined  algebraic
quantization approach  \c{Marolf}  for the case of continuous spectrum
of $\hat{\Lambda}_a$.  The  constraints  are  taken  into  account  as
follows. The  inner  product  of  the  wave  functions $\Phi_1(q)$ and
$\Phi_2(q)$ is introduced as
\beq
<\Phi_1, \Phi_2>  =
(\Phi_1,  \prod_{a=1}^M   \delta(\hat{\Lambda}_a) \Phi_2)
\equiv \int dq
\Phi_1^*(q)  \prod_{a=1}^M   \delta(\hat{\Lambda}_a) \Phi_2(q).
\l{2.3}
\eeq
Since the inner product \r{2.3} is degenerate,
the obtained  inner  product  space should be factorized:  states with
zero norm  are  set  to  be  equivalent  to  zero.  The  corresponding
factorspace should be completed in order to obtain a Hilbert space.

{\bf 3}. The nonabelian closed algebra case can be considered with the
help of    the    group    averaging    prescription    \c{GM}.    Let
$\tilde{\Lambda}_a$ be  Hermitian  quantum  constraints satisfying the
commutation relations
\bez
[\tilde{\Lambda}_a; \tilde{\Lambda}_b] = if^c_{ab} \tilde{\Lambda}_c,
\eez
where $f^c_{ab}$ are structure constants of the Lie algebra.  Consider
the corresponding representation of the Lie group of the form
\bez
T(\exp (i\mu^aL_a) ) = \exp (i\mu^a \tilde{\Lambda}_a),
\eez
where $L_a$ are generators of the Lie algebra with structure constants
$f^c_{ab}$, $\exp$ is  an  exponential  mapping  between  algebra  and
group. The inner product is introduced as follows \c{GM},
\beq
(\Phi_1, \int d_R g (det Ad g)^{-1/2} T(g) \Phi_2).
\l{2.4}
\eeq
Here $d_Rg$ is a  right-invariant  Haar  measure  on  the  Lie  group.
Eq.\r{2.3} is a partial case of \r{2.4}.

Note that  eq.\r{2.4}  can  be  rewritten  as  follows.  Consider  the
modified {\it non-Hermitian} constraints
\bez
\hat{\Lambda}_a = \tilde{\Lambda}_a - \frac{i}{2} f^b_{ab}
\eez
obeying the same commutation relations
\bez
[\hat{\Lambda}_a; \hat{\Lambda}_b] = if^c_{ab} \hat{\Lambda}_c,
\eez
Formula \r{2.4} will be rewritten then as
\beq
(\Phi_1, \int d_R g \exp (i\mu^a \hat{\Lambda}_a) \Phi_2).
\l{2.4m}
\eeq

{\bf 4.} The case of constrained algebra with structure functions
\beq
[\hat{\Lambda}_a; \hat{\Lambda}_b] = i \hat{\Lambda}_c \hat{U}^c_{ab},
\l{2.4c}
\eeq
where $\hat{U}^c_{ab}$, is more complicated \c{Sh}. One should use the
relationship between Dirac and BRST-BFV approaches and make use of the
BRST-BFV inner product \c{Marnelius}.   Introduce
additional Grassmannian    variables    $\overline{\Pi}_a$,    $\Pi^a$,
$a=\overline{1,M}$. The  quantum  constrained  system  with  an   open
algebra is specified by the B-charge
\bez
\hat{\Omega}_0 =  \sum_{n=1}^{\infty}
\hat{\Omega}^n{}_{a_1...a_n}^{b_1...b_{n-1}}
\overline{\Pi}_{b_1} ...
\overline{\Pi}_{b_{n-1}}
\frac{\partial}{\partial \overline{\Pi}_{a_1}} ...
\frac{\partial}{\partial \overline{\Pi}_{a_n}}
\eez
being a   Hermitian   nilpotent    operator,    $\hat{\Omega}_0^+    =
\hat{\Omega}_0$, $\hat{\Omega}_0   \hat{\Omega}_0   =  0$.  The  first
component of  the  B-charge  can  be  identified  with   the   quantum
constraint,
\bez
\hat{\Omega}_a^1 = \hat{\Lambda}_a.
\eez
The semiclassical structure of the B-charge should be as follows:
\beq
\hat{\Omega}_0 \sim   \Omega_0(\sqrt{h}   \hat{q},  \sqrt{h}  \hat{p},
\sqrt{h} \overline{\Pi},       \sqrt{h}       \frac{\partial}{\partial
\overline{\Pi}}).
\l{2.b}
\eeq
If the  quantum constraints depend on the small parameter according to
eq.\r{2.2}, the  higher  order  structure  functions  $\hat{\Omega}^n$
should depend on $h$ as follows,
\beq
\hat{\Omega}^n {}^{b_1...b_{n-1}}_{a_1...a_n} =
h^{n-2} \Omega^n
{}^{b_1...b_{n-1}}_{a_1...a_n} (\sqrt{h}\hat{q}, \sqrt{h} \hat{p})
+ h^{n-1} \Omega^{n (1)}
{}^{b_1...b_{n-1}}_{a_1...a_n} (\sqrt{h}\hat{q},  \sqrt{h}  \hat{p}) +
...
\l{2.4n}
\eeq
The classical  structure  functions  $\Omega^n$  were  constructed  in
\c{Henneaux} from the relation $\{ \Omega_0,  \Omega_0\} = 0$.  Higher
quantum corrections   $\Omega^{n(1)}$,   ...   can  be  calculated  in
analogous way. Let us investigate the corollaries of the conditions
$\hat{\Omega}_0^+ = \hat{\Omega}_0$ and $\hat{\Omega}_0 \hat{\Omega}_0
= 0$      for      the      operators      $\hat{\Lambda}_a$       and
$\hat{\Omega}^2{}^b_{a_1a_2}$. Property $\hat{\Omega}_0 \hat{\Omega}_0
= 0$ implies that
\bez
[\hat{\Lambda}_{a_1}; \hat{\Lambda}_{a_2}] + 2\hat{\Lambda}_b
\hat{\Omega}^2{}^b_{a_1a_2} = 0,
\eez
so that the operator $\hat{\Omega}^2{}^b_{a_1a_2}$ is related  to  the
structure functions  $\hat{U}^b_{a_1a_2}$  entering  to eq.\r{2.4c} as
follows,
\beq
\hat{\Omega}^2{}^b_{a_1a_2} = - \frac{i}{2}
\hat{U}^b_{a_1a_2}.
\eeq
We also see that the structure functions should depend on $h$ as
\bez
\hat{U}^b_{a_1a_2} = U^b_{a_1a_2} (\sqrt{h}\hat{q}, \sqrt{h} \hat{p}) +
h U^{(1) b}_{a_1a_2} (\sqrt{h}\hat{q}, \sqrt{h} \hat{p}) + ...
\eez
Since the Weyl symbol of the commutator is proportional to the Poisson
bracket of the operators (see Appendix A), in the leading order in $h$
one finds
\beq
\{ \Lambda_a; \Lambda_b \} = - \Lambda_c U^c_{ab}.
\l{2.5b}
\eeq

Property       $\hat{\Omega}_0^+      =
\hat{\Omega}_0$ implies that
\beq
({\Omega}^2{}^b_{a_1a_2})^* = -  {\Omega}^2{}^b_{a_1a_2}, \qquad
\Lambda_a^* = \Lambda_a,\qquad
\Lambda_a^{(1)*} - \Lambda_a^{(1)} =
2{\Omega}^2{}^b_{a_1a_2}.
\l{2.5c}
\eeq
We see  that  the  classical  constraints are real,  while the quantum
corrections $\Lambda_a^{(1)}$ have nontrivial imaginary part,
\beq
\Lambda_a^{(1)} = Re \Lambda_a^{(1)} + \frac{i}{2} U^d_{da}.
\l{2.a}
\eeq

The inner product of states is written as \c{Sh}
\beq
<\Phi_1, \Phi_2>    =    (\Phi_1,    \int     \prod_{a=1}^M     d\mu_a
d\overline{\Pi}_a d\Pi^a   e^{-\overline{\Pi}_a   \Pi^a   +   i  \mu_a
[\overline{\Pi}_a; \hat{\Omega}_0]_+} \Phi_2).
\l{2.5}
\eeq
Formula \r{2.5} coincides with \r{2.4m} for the closed-algebra case.

\subsection{"Elementary" semiclassical  states for constrained systems
and their inner product}

The most popular semiclassical approach to quantum  mechanics  is  the
WKB-approach based   on   substitution  of  rapidly  oscillating  wave
function to the Schrodinger equation and estimation of  the  accuracy.
However, there exist other types of wave functions which approximately
satisfy the Schrodinger  equation  (appendix  B).  Such  semiclassical
solutions can be somehow classified with the help of the Maslov theory
of Lagrangian manifolds with complex  germ  \c{Maslov2}.  It  happens,
however, that  one can consider first the wave packet solutions of the
Schrodinger equation such that uncertainties of coordinates and momenta
are of   orders   $O(\sqrt{h})$  or  after  rescaling  \r{2.1a}  $O(1)$
(contrary to $O(1/\sqrt{h})$ in the WKB-approach). Other semiclassical
wave functions (including WKB) can be viewed as superpositions of wave
packets. Thus,  wave packet states may be considered  as  "elementary"
semiclassical states,  while  other  semiclassical  wave functions are
"composed" states to be considered in section 6.

In the notations \r{2.1a},  elementary semiclassical state corresponds
to the wave function
\beq
\Psi(q) =  c  e^{\frac{i}{h}S}  e^{\frac{i}{h} P (q\sqrt{h} - Q)} f(q-
Q/\sqrt{h}).
\l{2.12}
\eeq
It is  specified  by  classical variables ($S\in {\bf R}$,  $P\in {\bf
R}^n$, $Q\in {\bf R}^n$) and "quantum" function $f$  which  is  smooth
and rapidly damps at the infinity.

Let us investigate the inner product of semiclassical states.

\subsubsection{Abelian case}

Formula \r{2.3} can be rewritten as
\beq
<\Phi,\Phi> =       \int       \prod_{a=1}^M       d\mu_a       (\Phi,
e^{i\mu_a\hat{\Lambda}_a} \Phi)
\l{2.13}
\eeq
It is necessary to calculate the wave function
\bez
\Phi^{\tau} = e^{-i\tau \mu_a \hat{\Lambda}_a} \Phi
\eez
as $h\to   0$   at   $\tau   =   -1$.   Note  that  it  satisfies  the
Schrodinger-type equation
\beq
i \frac{\partial \Phi^{\tau} }{\partial \tau } = \mu_a \hat{\Lambda}_a
\Phi^{\tau} = \left[
\frac{1}{h} \mu_a \Lambda_a (\sqrt{h}  \hat{q},  \sqrt{h}  \hat{p})  +
\mu_a \Lambda_a^{(1)} (\sqrt{h}  \hat{q},  \sqrt{h}  \hat{p}) + ...
\right] \Phi^{\tau}
\l{2.14}
\eeq
and initial  condition  \r{2.12}.  Let  us  look  for  the  asymptotic
solution in the following form
\beq
\Phi^{\tau}(q) = c e^{\frac{i}{h} S^{\tau} }  e^{\frac{i}{h}  P^{\tau}
(q\sqrt{h} - Q^{\tau})} f^{\tau} (q - \frac{Q^{\tau}}{\sqrt{h}})
\l{2.15}
\eeq
Substituting expression \r{2.15} to eq.\r{2.14}, we find:
\beb
\left[
- \frac{1}{h} (\dot{S}^{\tau} - P^{\tau} \dot{Q}^{\tau}) -
\frac{1}{\sqrt{h}} \dot{P}^{\tau}  \xi  +  i  \frac{\partial}{\partial
\tau} - \frac{i}{\sqrt{h}} \dot{Q}^{\tau} \frac{\partial}{\partial \xi}
\right] f^{\tau}(\xi)
= \\
\frac{1}{h} \mu_a \Lambda_a(Q^{\tau} +  \sqrt{h}  \xi,  P^{\tau}  -  i
\sqrt{h} \frac{\partial}{\partial \xi}) f^{\tau}(\xi)
+ \mu_a \Lambda_a^{(1)} (Q^{\tau} +  \sqrt{h}  \xi,  P^{\tau}  -  i
\sqrt{h} \frac{\partial}{\partial \xi}) f^{\tau}(\xi) + ...
\l{2.15a}
\eeb
where $\xi = q- Q^{\tau}/\sqrt{h}$. It is shown in Appendix A that the
operator $\Lambda_a(Q     +    \sqrt{h}    \xi,    P    -    i\sqrt{h}
\frac{\partial}{\partial \xi})$ is expanded in $\sqrt{h}$ as
\bez
\Lambda_a(Q     +    \sqrt{h}    \xi,    P    -    i\sqrt{h}
\frac{\partial}{\partial \xi})   =   \Lambda_a(Q,P)  +  \sqrt{h}  (\Xi
\Lambda_a) (Q,P) + \frac{h}{2} (\Xi^2 \Lambda_a)(Q,P) + ...
\eez
where
\bez
\Xi =    \xi     \frac{\partial}{\partial     Q}     +     \frac{1}{i}
\frac{\partial}{\partial \xi} \frac{\partial}{\partial P}.
\eez
The terms of the order $O(h^{-1})$ in eq.\r{2.15a} give us an equation
on the phase factor $S^{\tau}$
\beq
\dot{S}^{\tau} =      P^{\tau}      \dot{Q}^{\tau}       -       \mu_a
\Lambda_a(Q^{\tau},P^{\tau})
\l{2.16}
\eeq
We see that $S^{\tau}$ is the action on the classical trajectory.

The terms of the order $O(h^{-1/2})$ lead to classical equations
\beq
\dot{Q}^{\tau} =   \mu_a   \frac{\partial    \Lambda_a}{\partial    P}
(Q^{\tau},P^{\tau}),
\qquad
\dot{P}^{\tau} =  - \mu_a   \frac{\partial    \Lambda_a}{\partial    Q}
(Q^{\tau},P^{\tau}),
\l{2.17}
\eeq
Under conditions  \r{2.16} and \r{2.17},  we find in the leading order
in $h$ the following equation on $f$
\beq
i \frac{\partial f^{\tau}(\xi)}{\partial \tau} = \left[
\frac{1}{2}\mu_a (\Xi^2    \Lambda_a)(Q^{\tau},P^{\tau})    +    \mu_a
\Lambda_a^{(1)} (Q^{\tau},P^{\tau})
\right] f^{\tau}(\xi)
\l{2.17m}
\eeq
with the quadratic Hamiltonian.

Let us substitute the wave function \r{2.15} at $\tau = -1$ to formula
\r{2.13}. First    of    all,    notice   that   the   inner   product
$(\Phi,\Phi^{\tau})$ is not exponentially small only if
\beq
|P^{\tau} - P| \le O(\sqrt{h}),
\qquad
|Q^{\tau} - Q| \le O(\sqrt{h}).
\l{2.18}
\eeq
Namely, the  wave function $\Phi^{\tau}(q)$ is not exponentially small
only if   $q-Q^{\tau}/\sqrt{h}   =   O(1)$,   so    that    $\Phi^*(q)
\Phi^{\tau}(q)$ will be not exponentially small if
\bez
q-Q^{\tau}/\sqrt{h} = O(1),
\qquad
q-Q/\sqrt{h} = O(1).
\eez
Therefore, $|Q^{\tau} -Q|$ should be $\le O(\sqrt{h})$.  If $|P^{\tau}
- P| > O(\sqrt{h})$,  the integral $\int dq \Phi^*(q)  \Phi^{\tau}(q)$
will contain  rapidly  oscillating  factor  and  be then exponentially
small.

Several cases should be considered.  Here we investigate the  simplest
"general  position"  or  "free" case (taking place in QED,  Yang-Mills
theories),  when the action of the gauge group on the classical  phase
space is free,  i.e.  the stationary subgroup of any point is trivial.
This means that $Q^{\tau}  =Q$,  $P^{\tau}  =  P$  only  if  $\tau=0$.
Nonfree case will be briefly discussed in section 7.

Conditions \r{2.18} are satisfied in the free case
only if $\tau = O(\sqrt{h})$.  It is  convenient  then  to  perform  a
substitution
\bez
\mu_a \Rightarrow \mu_a \sqrt{h}
\eez
for eq.\r{2.13}.  After  substitution $q - Q^0/\sqrt{h} = \xi$ formula
\r{2.13} is taken to the form
\beq
<\Phi, \Phi> = h^{M/2} |c|^2 \int d\mu d\xi e^{\frac{i}{h}
(S^{-\sqrt{h}} - S^0 + P^{-\sqrt{h}} (Q^0 - Q^{-\sqrt{h}}) )}
e^{\frac{i}{\sqrt{h}} (P^{-\sqrt{h}} - P^0)\xi }
t^{0*} (\xi) f^{-\sqrt{h}} (\xi - \frac{Q^{-\sqrt{h}} - Q^0}{\sqrt{h}})
\l{2.19}
\eeq
Since
\bez
S^{-\sqrt{h}} - S^0 + P^{-\sqrt{h}} (Q^0 - Q^{-\sqrt{h}})
= -\sqrt{h}  (\dot{S}^0  -  P^0 \dot{Q}^0) + \frac{h}{2} (\ddot{S}^0 -
P^0 \ddot{Q}^0) - h\dot{P}^0 \dot{Q}^0 + o(h),
\eez
expression \r{2.19} is taken to the form
\beq
<\Phi,\Phi> \simeq
h^{M/2} |c|^2 \int d\mu d\xi
e^{ - \frac{i}{\sqrt{h}}
(\dot{S}^0  -  P^0 \dot{Q}^0)}
e^{ \frac{i}{2} (\ddot{S}^0 - P^0 \ddot{Q}^0 - 2\dot{P}^0 \dot{Q}^0)}
e^{-i\dot{P}^0\xi} f^*(\xi) f(\xi + \dot{Q}^0).
\l{2.20}
\eeq
We see that if
\bez
\dot{S}^0 - P^0 \dot{Q}^0 = - \mu_a \Lambda_a (Q^0,P^0) \ne 0,
\eez
the integrand entering to eq.\r{2.20} contains a  rapidly  oscillating
function. Therefore,  integral  \r{2.20} is exponentially small.  We
see that the wave function  \r{2.12}  $\Phi$  is  a  {\it  nontrivial}
semiclassical state, $<\Phi, \Phi> \ne 0$, only if
\beq
\Lambda_a(Q,P) = 0.
\l{2.21}
\eeq
This means that the classical state should belong  to  the  constraint
surface.

Differentiating eq.\r{2.16}  with respect to $\tau$ at $\tau = 0$,  we
find
\bez
\ddot{S}^0 -    P^0    \ddot{Q}^0    =    \dot{P}^0    \dot{Q}^0     -
\frac{d}{d\tau}|_{\tau=0} \mu_a     \Lambda_a(Q^{\tau},P^{\tau})     =
\dot{P}^0 \dot{Q}^0.
\eez
Making use of the  Baker-Hausdorff  formula,  we  simplify  eq.\r{2.20}
under condition \r{2.21},
\bez
<\Phi,\Phi> \simeq   h^{M/2}  |c|^2  \int  d\mu  d\xi  f^*(\xi)  e^{-i
(\dot{P}^0 \xi - \dot{Q}^0 \frac{1}{i} \frac{\partial}{\partial\xi})  }
f(\xi).
\eez
It follows  from Hamiltonian equations \r{2.17} that the inner product
of semiclassical states is
\beq
<\Phi,\Phi> \simeq h^{M/2} |c|^2 \int d\xi d\mu f^*(\xi)
(e^{
i\mu_a (\frac{\partial \Lambda_a}{\partial Q} \xi +
\frac{\partial \Lambda_a}{\partial            P}           \frac{1}{i}
\frac{\partial}{\partial \xi} )}f)(\xi)
= h^{M/2} |c|^2 (f, \prod_{a=1}^M (2\pi \delta(\Xi \Lambda_a) )f)
\l{2.22}
\eeq
under condition \r{2.21}.

We see that the normalization factor $|c|$  should  be  of  the  order
$h^{-M/4}$ in  order  to  make the norm of state \r{2.12} to be of the
order $O(1)$.  We also notice that the semiclassical inner product  is
obtained from  the quantum formula \r{2.3} by linearization procedure:
the constraint operators
\bez
\Lambda_a (Q+    \sqrt{h}    \xi,    P    +    \frac{1}{i}    \sqrt{h}
\frac{\partial}{\partial \xi})
\eez
should be  linearized.  It  has  been  understood  that  an  important
condition of validity of linearization prescription is free action of
classical gauge group.

\subsubsection{Nonabelian and open-algebra case}

Let us investigate the inner product \r{2.4m} for the nonabelian gauge
group in the semiclassical approximation.  The  free
case only  will  be  considered.  The  integral  in  the inner product
\r{2.4m} is not exponentially small only if $\mu = O(\sqrt{h})$. After
rescaling $\mu \Rightarrow \mu\sqrt{h}$ formula \r{2.4m} takes the form
\bez
h^{M/2} \int d\mu J(\mu\sqrt{h}) (\Phi^0,\Phi^{-\sqrt{h}}),
\eez
where $\Phi^{\tau}$   is   a   solution  of  the  Cauchy  problem  for
eq.\r{2.14} with initial  condition  $\Phi^0  =  \Phi$  \r{2.12}.  The
jacobian $J(\mu)$ normalized as $J(0)=1$ is defined from formula $d_Rg
= d\mu J(\mu)$.  Analogously to the previous subsubsection,  we obtain
condition \r{2.21} and formula \r{2.22}.

The open-algebra  case  can be investigated in the same way.  Consider
the wave function
\beq
\Phi^{\tau} (q,\Pi,  \overline{\Pi}) = \exp  \{\tau  \overline{\Pi}_a
\Pi^a - i\tau \mu_a [\overline{\Pi}_a, \hat{\Omega}_0]_+ \} \Phi(q).
\l{2.i1}
\eeq
obeying the equation
\bez
i \frac{\partial   \Phi^{\tau}   }{   \partial   \tau   }   =   [\mu_a
(\hat{\Lambda}_a +      \sum_{n=2}^{\infty}      n      \hat{\Omega}^n
{}_{a_1...a_{n-1}a}^{b_1...b_{n-1}} \overline{\Pi}_{b_1}
... \overline{\Pi}_{b_{n-1}}
\frac{\partial}{\partial \overline{\Pi}_{a_1}}
...
\frac{\partial}{\partial \overline{\Pi}_{a_n}})  +  i  \overline{\Pi}_a
\Pi^a] \Phi^{\tau}
\eez
and initial condition \r{2.12}.  Let us look for the solution  of  the
Cauchy problem in the form
\bez
\Phi^{\tau}(q,\Pi,\overline{\Pi}) = c e^{\frac{i}{h}S^{\tau}}
e^{\frac{i}{h} P^{\tau}   (q\sqrt{h}-    Q^{\tau})}    f^{\tau}    (q-
\frac{Q^{\tau}}{\sqrt{h}}, \Pi, \overline{\Pi}).
\eez
Take into  account  that  terms with $\hat{\Omega}^n$ are of the order
$O(h^{n-2})$ and can therefore be neglected at $n\ge  3$.  Analogously
to the  previous  subsection,  we  obtain  eq.\r{2.16} for $S^{\tau}$,
eq.\r{2.17} for $P^{\tau}$,  $Q^{\tau}$ and the following equation for
$f^{\tau}$ which differs from \r{2.17m},
\beq
i \frac{\partial f^{\tau}}{\partial \tau} = \left[
\frac{1}{2} \mu_a   (\Xi^2   \Lambda_a)(Q^{\tau},P^{\tau})   +   \mu_a
\Lambda_a^{(1)}(Q^{\tau},P^{\tau}) + i \overline{\Pi}_a \Pi^a + 2\mu_a
\Omega^2{}^{b_1}_{a_1a} (Q^{\tau},P^{\tau}) \overline{\Pi}_{b_1}
\frac{\partial}{\partial \overline{\Pi}_{a_1}}
\right] f^{\tau}.
\l{2.i2}
\eeq
The integrand in the inner product \r{2.5} is not exponentially  small
only if   $\mu   =  O(\sqrt{h})$.  After  rescaling  $\mu  \Rightarrow
\mu\sqrt{h}$, we obtain conditions \r{2.21}.  The inner product  takes
the form
\bez
<\Phi,\Phi> \simeq    h^{M/2}    |c|^2   \int   \prod_{a=1}^M   d\mu_a
d\overline{\Pi}_a d\Pi^a (f, e^{-\overline{\Pi}_a \Pi^a - i (\dot{P}^0
\xi - \dot{Q}^0 \frac{1}{i} \frac{\partial}{\partial \xi})} f).
\eez
We also obtain formula \r{2.22}.

We see  that  the  obtained  expression  for  the  semiclassical inner
product is valid for the nonabelian and open algebra  cases.  We  also
see that  the  nontriviality  of  condition \r{2.21} is also valid for
such cases.

\subsection{Semiclassical bundle}

We see that a semiclassical wave function \r{2.12} is specified if:

(i) a   set   $X  =  (S,P,Q)$  ("classical  state")  satisfying  the
requirement \r{2.21} is specified;

(ii) a smooth rapidly damping at the infinity function  $f\in  {\cal
S}({\bf R}^n)$ is specified.

Semiclassical wave  functions  will  be denoted as $(X,f)$.  The inner
product of semiclassical wave functions is introduced as
\beq
<(X,f_1);(X,f_2)> =     (f_1,
\prod_{a=1}^M      2\pi      \delta
((\Xi\Lambda_a)(X)) f_2),
\l{2.23}
\eeq
provided that classical gauge group is free.  Note that  according  to
Appendix A  the  operators  $\Xi\Lambda_a$  commute  each other on the
constraint surface \r{2.21}, since
\bez
[\Xi\Lambda_a, \Xi\Lambda_b]     =     -i\{\Lambda_a,\Lambda_b\}     =
iU_{ab}^c\Lambda_c = 0.
\eez

By ${\cal F}_X^0$ we denote the inner product space of complex functions
$f \in {\cal S}({\bf R}^n)$ with the inner product \r{2.23}.  Since it
is degenerate,  one should  factorize  the  corresponding  pre-Hilbert
space as follows: two functions $f_1$ and $f_2$ are called equivalent,
$f_1 \equiv f_2$, if their difference $\phi = f_1-f_2$ has zero norm,
\bez
(\phi,
\prod_{a=1}^M      2\pi      \delta
((\Xi\Lambda_a)(X)) \phi) = 0.
\eez
For example, consider the wave function $\phi$ of the form
\bez
\phi = (\Xi \Lambda_a) (X) \chi^a.
\eez
It has zero norm, so that the transformation
\beq
f \to f + (\Xi \Lambda_a) (X) \chi^a
\l{2.24}
\eeq
takes the  semiclassical  wave  function  $(X,f)$  to  the  equivalent
semiclassical wave function $(X,f  +  (\Xi  \Lambda_a)  (X)  \chi^a)$.
Since the  classical state $X$ does not vary during the transformation
\r{2.24}, it can be called as a "small" gauge  transformation  of  the
semiclassical wave function. "Large" gauge transformations varying the
classical state $X$ will be considered in the next section.

By ${\cal F}_X = {\cal  F}_X^0/\sim$  we  denote  the  factorspace  of
equivalence classes $[f]$. Let the Hilbert space
$\overline{\cal F}_X$ be a completeness of the pre-Hilbert space
${\cal F}_X$.

We see  that it is more correct to consider "elementary" semiclassical
states as pairs $(X,\overline{f})$,  $\overline{f} \in  \overline{\cal
F}_X$ rather   that   pairs  $(X,f)$.  The  set  of  all  'elementary"
semiclassical states can be viewed as a bundle. The base of the bundle
is ${\cal X} = \{ (S,P,Q)|  \Lambda_a(Q,P) = 0\}$, the fibres are Hilbert
spaces $\overline{\cal  F}_X$.  Such  a   bundle   was   called   {\it
semiclassical} in  \c{Sh1,Sh2}.  "Elementary" semiclassical states are
then points on the semiclassical bundle.

\section{Gauge equivalent semiclassical states}

\subsection{"Small" and "large" gauge transformations}

Property of gauge  invariance  plays  an  important  role  in
classical mechanics  of constrained systems.  This property means that
classical constraints  generate  gauge  transformations  on  classical
phase space.  Classical  states  that  belong  to  one  orbit of gauge
transformation are called equivalent.

An analogous property takes place  for  the  semiclassical  theory  as
well. Namely, the state $\hat{\Lambda}_a X^a$ has zero norm \c{GM,Sh}.
Let
\bez
X^a(q) = e^{\frac{i}{h}S} e^{\frac{i}{h} P (q\sqrt{h}-Q)}  \chi^a(q  -
Q/\sqrt{h}).
\eez
The wave  function  $h^{-1/2}  \hat{\Lambda}_a  X^a$ has the following
form in the leading order in $\sqrt{h}$:
\bez
h^{-1/2}  \hat{\Lambda}_a  X^a =
e^{\frac{i}{h}S} e^{\frac{i}{h} P (q\sqrt{h}-Q)} (\Xi \Lambda_a \chi^a)
(q  - Q/\sqrt{h}),
\eez
since $\Lambda_a(Q,P)=0$.  We  have   obtained   the   "small"   gauge
transformation \r{2.24}.

To obtain a "large" gauge transformation,  note that semiclassical wave
functions
\bez
\Phi \qquad and \qquad e^{-i\tau \mu_a \hat{\Lambda}_a} \Phi
\eez
should be  called  gauge-equivalent.  However,   the   wave   function
$\Phi^{\tau} = e^{-i\tau \mu_a \hat{\Lambda}_a} \Phi$ has been already
calculated in  the  semiclassical  approximation.  It  has  the   form
\r{2.15}, where    $S^{\tau}$   satisfies   eq.\r{2.16},   $P^{\tau}$,
$Q^{\tau}$ obey the classical Hamiltonian  equations  \r{2.17},  while
$f^{\tau}$ is a solution of eq.\r{2.17m}.

By $\lambda_{\mu\tau}$  we  denote  the mapping taking $X= (S,P,Q)$ to
$X^{\tau} = (S^{\tau},P^{\tau},Q^{\tau})$ which is a  classical  gauge
transformation. By $V^0(\lambda_{\mu\tau} X\gets X) : {\cal F}_X^0 \to
{\cal F}_{\lambda_{\mu\tau} X}^0$ we denote the  operator  taking  the
initial condition  for the Cauchy problem \r{2.17m} to the solution of
the Cauchy problem,
\bez
f^{\tau} = V^0(\lambda_{\mu\tau}X \gets X) f^0.
\eez
The semiclassical wave functions
\bez
(X,f) \qquad and \qquad
(\lambda_{\mu\tau}X, V^0(\lambda_{\mu\tau}X \gets X) f)
\eez
are gauge-equivalent  then.  This  is  a "large" gauge transformation.
Obviously, it  conserves  the   conditions   $\Lambda_c(X)=0$,   since
$\{\Lambda_a; \Lambda_c\} = 0$ on the constraint surface.

\subsection{Unitarity problem}

Let us  show  that  semiclassical  gauge  transformation conserves the
inner product  \r{2.23}.  First,  consider  the   commutator   between
operators
$i\frac{\partial}{\partial \tau} - \mu_a \left[
\frac{1}{2} (\Xi^2   \Lambda_a)(Q^{\tau},P^{\tau})  +  \Lambda_a^{(1)}
(Q^{\tau},P^{\tau})
\right]$ and   $\Xi\Lambda_b   (Q^{\tau},P^{\tau})$.   Making  use  of
results of Appendix A, we find that it can be presented as
\beq
\left[ i\frac{\partial}{\partial \tau} - \mu_a \left[
\frac{1}{2} (\Xi^2   \Lambda_a)(Q^{\tau},P^{\tau})  +  \Lambda_a^{(1)}
(Q^{\tau},P^{\tau})
\right];
\Xi\Lambda_b   (Q^{\tau},P^{\tau}) \right]
= i (\Xi \{ \mu_a \Lambda_a, \Lambda_b \}) (Q^{\tau},P^{\tau}).
\l{3.1}
\eeq
It follows from eq.\r{2.5b} that
\beq
i \Xi \{\mu_a\Lambda_a; \Lambda_b\} = - i \Xi (\Lambda_c U^c_{ab}) = -
i U^c_{ab} \Xi \Lambda_c
\l{3.5a}
\eeq
on the constraint surface $\Lambda_c=0$.  Thus, the commutator \r{3.1}
takes the form
\beq
\left[ i\frac{\partial}{\partial \tau} - \mu_a \left[
\frac{1}{2} (\Xi^2   \Lambda_a)(Q^{\tau},P^{\tau})  +  \Lambda_a^{(1)}
(Q^{\tau},P^{\tau})
\right];
\Xi\Lambda_b   (Q^{\tau},P^{\tau}) \right]
= -i U^c_{ab} (Q^{\tau},P^{\tau}) (Xi \Lambda_c) (Q^{\tau},P^{\tau}).
\l{3.2}
\eeq
Let $f_1^{\tau}$ and $f_2^{\tau}$ be solutions  of  eq.\r{2.17m}.  The
time derivative  of  their  inner  product $<f_1^{\tau},  f_2^{\tau}>$
\r{2.23} can be written as
\beb
i \frac{\partial}{\partial \tau} <f_1^{\tau}, f_2^{\tau}> =
(-i \dot{f}_1^{\tau},  \prod_a  (2\pi  \delta(\Xi \Lambda_a)(Q^{\tau},
P^{\tau})) f_2^{\tau})
+
({f}_1^{\tau},  \prod_a  (2\pi  \delta(\Xi \Lambda_a)(Q^{\tau},
P^{\tau})) i\dot{f}_2^{\tau})
\\
+
({f}_1^{\tau},  i \frac{\partial}{\partial \tau}\left\{\prod_a
(2\pi  \delta(\Xi \Lambda_a)(Q^{\tau},
P^{\tau}))\right\} f_2^{\tau})
\l{3.3}
\eeb
Making use  of  equation  of  motion  \r{2.17}.  Taking  into  account
relation \r{2.a}, we take expression \r{3.3} to the form
\beq
i \frac{\partial}{\partial \tau} <f_1^{\tau}, f_2^{\tau}> =
i\mu_a U^b_{ba}(Q^{\tau},P^{\tau} <f_1^{\tau}, f_2^{\tau}> +
(f_1^{\tau},
\left[
i\frac{\partial}{\partial \tau} - \mu_a
\frac{1}{2} (\Xi^2   \Lambda_a)(Q^{\tau},P^{\tau}),
\int d\rho e^{i\rho_b (\Xi\Lambda_b)(Q^{\tau},P^{\tau})}
\right]
f_2^{\tau})
\l{3.4}
\eeq
Commutation relation \r{3.2} implies the following commutation rule
\bey
e^{-i\rho_b (\Xi\Lambda_b)(Q^{\tau},P^{\tau})}
\left\{
i\frac{\partial}{\partial \tau} - \mu_a
\frac{1}{2} (\Xi^2   \Lambda_a)(Q^{\tau},P^{\tau})
\right\}
e^{i\rho_b (\Xi\Lambda_b)(Q^{\tau},P^{\tau})}
=
i\frac{\partial}{\partial \tau} - \mu_a
\frac{1}{2} (\Xi^2   \Lambda_a)(Q^{\tau},P^{\tau})
\\
- i \rho_b \left[
(\Xi\Lambda_b)(Q^{\tau},P^{\tau});
i\frac{\partial}{\partial \tau} - \mu_a
\frac{1}{2} (\Xi^2   \Lambda_a)(Q^{\tau},P^{\tau})
\right]
\eey
Higher order terms will vanish since $[\Xi\Lambda_a; \Xi\Lambda_b] = -
i \{\Lambda_a, \Lambda_b\} = 0$ on the constraint surface. Therefore,
\bez
\left[
i\frac{\partial}{\partial \tau} - \mu_a
\frac{1}{2} (\Xi^2   \Lambda_a)(Q^{\tau},P^{\tau});
e^{i\rho_b (\Xi\Lambda_b)(Q^{\tau},P^{\tau})}
\right]
= - i\mu_a U^c_{ab}(Q^{\tau},P^{\tau}) \rho_b \frac{\partial}{\partial
\rho_c}
e^{i\rho_d (\Xi\Lambda_d)(Q^{\tau},P^{\tau})}
\eez
Integrating this expression over $\rho$ by parts, we find that
\bez
\left[
i\frac{\partial}{\partial \tau} - \mu_a
\frac{1}{2} (\Xi^2   \Lambda_a)(Q^{\tau},P^{\tau});
\int d\rho
e^{i\rho_b (\Xi\Lambda_b)(Q^{\tau},P^{\tau})}
\right]
= i\mu_a U^b_{ab}(Q^{\tau},P^{\tau})
\int d\rho e^{i\rho_d (\Xi\Lambda_d)(Q^{\tau},P^{\tau})}
\eez
Substituting this result to eq.\r{3.4},  we obtain that "large"  gauge
transformations conserve the inner product,
\bez
<f_1^{\tau}, f_2^{\tau}> = const.
\eez
This implies   that   zero-norm  semiclassical  states  are  taken  to
zero-norm states. Therefore, one can correctly define the operators
$V(\lambda_{\mu\tau} X\gets X) : {\cal F}_X^0/\sim \equiv
{\cal F}_X \to
{\cal F}_{\lambda_{\mu\tau} X}^0/\sim
\equiv {\cal F}_{\lambda_{\mu\tau} X}$ by the formula
\bez
V(\lambda_{\mu\tau} X\gets X) [f] =
[V^0(\lambda_{\mu\tau} X\gets X) f].
\eez
The operators  $V(\lambda_{\mu\tau} X\gets X)$ are also unitary.  They
can be uniquely extended to the completion
$\overline{\cal F}_{X}$,
\bez
\overline{V}(\lambda_{\mu\tau} X\gets X) : \overline{\cal F}_X \to
\overline{\cal F}_{\lambda_{\mu\tau} X}.
\eez
Thus, the semiclassical states
\bez
(X,\overline{f}) \qquad and \qquad
(\lambda_{\mu\tau}X,\overline{V}(\lambda_{\mu\tau}X \gets           X)
\overline{f})
\eez
are equivalent.  Gauge transformations appears to be morphisms  of  the
semiclassical bundle.

\subsection{Quasigroup properties}

Let us show that composition of gauge equivalence transformations is a
gauge equivalence transformation,  i.e.  for any $X \in {\cal X}$  and
sufficiently small $\mu_1$,  $\mu_2$ there exist $\mu_3(\mu_1,\mu_2,X)$
such that
\bez
(\lambda_{\mu_1} \lambda_{\mu_2} X,
\overline{V}_{\mu_1}(\lambda_{\mu_1} \lambda_{\mu_2}      X      \gets
\lambda_{\mu_2} X)
\overline{V}_{\mu_2}(\lambda_{\mu_2} X \gets X) \overline{f} ) =
(\lambda_{\mu_3} X,
\overline{V}_{\mu_3}(\lambda_{\mu_3} X \gets X) \overline{f} ), \qquad
\overline{f} \in \overline{\cal F}_X.
\eez
This means that set of gauge transformations form a local  Batalin
quasigroup \c{Batalin}.

First of all, investigate the classical gauge transformations.

{\bf 1.}  Introduce  the first-order differential operators $\delta_a$
from the relations:
\beq
\mu^a (\delta_a f) (X) = \frac{d}{d\tau}|_{\tau=0} f(\lambda_{\mu\tau}
X).
\l{3.5}
\eeq
It follows from definition of $\lambda_{\mu\tau}$ that
\bez
\delta_a =      P      \frac{\partial      \Lambda_a}{\partial      P}
\frac{\partial}{\partial S} + \frac{\partial\Lambda_a}{\partial P}
\frac{\partial}{\partial Q} - \frac{\partial \Lambda_a}{\partial Q}
\frac{\partial}{\partial P}
\eez
The operators $\delta_a$ satisfy the following commutation relations
\bez
[\delta_a; \delta_b]  =  P  \frac{\partial}{\partial  P}  \{\Lambda_a;
\Lambda_b\} \frac{\partial}{\partial S} +
\frac{\partial  \{\Lambda_a;
\Lambda_b \}}{\partial P} \frac{\partial}{\partial Q}
- \frac{\partial  \{\Lambda_a;
\Lambda_b \}}{\partial Q} \frac{\partial}{\partial P}
\eez
since $\{\Lambda_a;  \Lambda_b\} = 0$ on the  constraint  surface.  It
follows from eq.\r{2.5b} that
\bez
[\delta_a; \delta_b] = -U^c_{ab}(Q,P) \delta_c.
\eez
This means  that  the operators $\delta_a$ form a Batalin quasialgebra
\c{Batalin}.

It is shown in \c{Batalin} that the quasialgebra property implies  the
quasigroup property: for  all  $X\in  {\cal  X}$   and
sufficiently small $\mu_1$, $\mu_2$
\beq
\lambda_{\mu_1} \lambda_{\mu_2} X = \lambda_{\mu_3(\mu_1,\mu_2,X)}X
\l{3.4a}
\eeq
for some $\mu_3$.

{\bf 2.} Let us justify the formula
\beq
\overline{V}_{\mu_1} (\lambda_{\mu_1}\lambda_{\mu_2}      X      \gets
\lambda_{\mu_2} X)
\overline{V}_{\mu_2} (\lambda_{\mu_2}  X \gets  X)
= \overline{V}_{\mu_3} (\lambda_{\mu_3}  X \gets  X),
\l{3.12}
\eeq
provided that $\lambda_{\mu_1} \lambda_{\mu_2}  X=  \lambda_{\mu_3}X$.
Remind that  the  operator  $V_{\mu\tau} (\lambda_{\mu\tau} X \gets X)
\equiv V_{\mu\tau}[X]$ is defined as follows.  Let $V^0_{\mu\tau} [X]$
be the  operator  taking  initial condition for the cauchy problem for
equation
\beq
i \frac{\partial      f^{\tau}      }{\partial\tau}      =      \mu^a
H_a(\lambda_{\mu\tau} X) f^{\tau}
\l{3.12a}
\eeq
to the solution of the Cauchy problem. Here
\bez
H_a(Q,P) = \frac{1}{2} \Xi^2 \Lambda_a (Q,P) + \Lambda_a^{(1)} (Q,P).
\eez
Since the operator $V_{\mu\tau}^0[X]$ conserves  the  norm  and  takes
zero-norm states to zero-norm states, one can consider the operator
$V_{\mu\tau}: {\cal F}^0_X/\sim  \to  {\cal  F}^0_{\lambda_{\mu\tau}X}
/\sim$ being also isometric. It can be extended to
$\overline{{\cal F}^0_X/\sim}$: $\overline{V}_{\mu\tau}:
\overline{\cal F}_X  \to  \overline{\cal  F}_{\lambda_{\mu\tau}X}$.

{\bf 3.}  First  of  all,  investigate  corollaries of eq.\r{3.12} for
infinitesimal operators $H_a(X)$.

Let us  consider  the  composition  of  transformation  $\lambda_{\mu}
\lambda_{\nu\tau} \lambda_{-\mu}$.  For  some function $\rho(\tau,X)$,
one locally has
\bez
\lambda_{\mu} \lambda_{\nu\tau} \lambda_{-\mu} X
= \lambda_{\rho(\tau,X)} X.
\eez
Denote
\bez
(W_{\mu} f)(X) = f(\lambda_{\mu}X).
\eez
One has then
\bez
\frac{d}{dt}|_{t=0}
(W_{-\mu} W_{\nu\tau} W_{\mu} F)(X) = \frac{d}{d\tau}|_{\tau=0}
(W_{\rho(\tau,X)} f)(X).
\eez
In the leading order in $\tau$, one has
\bez
\rho(\tau,X) \sim \overline{\nu}^a \tau
\eez
for some $\overline{\nu}^a$. It follows from definition \r{3.5} of the
operator $\delta_a$ that
\beq
W_{-\mu} \nu^a \delta_a W_{\mu} = \overline{\nu}^b \delta_b.
\l{i.1}
\eeq
where $\overline{\nu}^b$ linearly depends on $\nu^a$. Denote the
corresponding matrix of linear transformation as $(Ad_X \mu)_a^b$, so that
\beq
\overline{\nu}^b = (Ad_X \mu)_a^b \nu^a
\l{i.2}
\eeq
and
\bez
\rho(\tau,X) \sim (Ad_X \mu)_a^b \nu^a \tau.
\eez
Denote $\lambda_{-\mu}X = Y$.

If the property \r{3.12} is satisfied, then
\beb
< f,     V^0_{\mu}     (\lambda_{\mu}\lambda_{\nu\tau}     Y     \gets
\lambda_{\nu\tau} Y) V^0_{\nu\tau}(\lambda_{\nu\tau} Y \gets Y) g> =\\
< f,        V^0_{\rho(\tau,\lambda_{\mu}Y)}       (\lambda_{\rho(\tau,
\lambda_{\mu} Y)} \lambda_{\mu} Y  \gets  \lambda_{\mu}  Y)  V_{\mu}^0
(\lambda_{\mu} Y \gets Y) g>
\l{3.13}
\eeb
for $f\in {\cal F}^0_{\lambda_{\mu} \lambda_{\nu\tau} Y}$, $g\in {\cal
F}_Y^0$. Consider  this identity as $\tau\to 0$.  In the leading order
in $\tau$,  this relation is trivial.  Consider the  first  nontrivial
order. One has:
\bey
V_{\mu}^0(\lambda_{\mu} \lambda_{\nu\tau} Y \gets \lambda_{\nu\tau} Y)
\sim V_{\mu}[Y] + \tau (\nu^a \delta_a V_{\mu})[Y];\\
V_{\nu\tau}^0(\lambda_{\nu\tau} Y  \gets  Y)  \sim  1  -  i\tau  \nu^a
H_a(Y);\\
V^0_{\rho(\tau,\lambda_{\mu}Y)}    (\lambda_{\rho(\tau,\lambda_{\mu}Y)}
\lambda_{\mu} Y  \gets  \lambda_{\mu}  Y)  \sim  1   -   i\tau   \nu^a
(Ad_{\lambda_{\mu}Y} \mu)_a^b H_b(\lambda_{\mu}Y).
\eey
Combining the terms of the order $O(\tau)$ in eq.\r{3.13}, one obtains:
\beq
<f, ((\delta_a V_{\mu}^0)[Y] - i V_{\mu}^0[Y] H_a(Y)) g >
= -   i   (Ad_{\lambda_{\mu}Y}   \mu)_a^b   <f,    H_b(\lambda_{\mu}Y)
V^0_{\mu}[Y] g>.
\l{3.14}
\eeq
Property \r{3.14} is very important.

{\bf 4.} Consider the substitution $\mu \rightarrow  \mu  t$  and  let
$t\to 0$.  First,  obtain  an equation for $(Ad_{\lambda_{\mu t}Y} \mu
t)_a^b$. Relations \r{i.1} and \r{i.2} can be presented as
\bez
W_{\mu t} \delta_a W_{-\mu t} = (Ad_X (-\mu t))_a^b \delta_b.
\eez
It follows from definition \r{3.5} that
\bez
\frac{d}{dt} W_{\mu t} = \mu^b \delta_b W_{\mu t} =
W_{\mu t}\mu^b \delta_b.
\eez
Therefore,
\bey
\frac{d}{dt} (Ad_x (-\mu t))_a^b \delta_b = W_{\mu t} [\mu^b \delta_b;
\delta_a] W_{-\mu t} = \mu^b U^c_{ab} (\lambda_{\mu t}  X)  W_{\mu  t}
\delta_c W_{-\mu  t}  =  \mu^b  U^c_{ab}(\lambda_{\mu t}X) (Ad_X (-\mu
t))_c^d \delta_d
\eey
and
\beq
\frac{d}{dt} (Ad_X (-\mu t))_a^d = \mu^b U^c_{ab}(\lambda_{\mu  t}  X)
(Ad_X (-\mu t))_c^d
\l{3.8}
\eeq
(cf. \c{Batalin}).

Making use of eqs.\r{i.1} and \r{i.2} twice, one finds
\bez
\delta_a = W_{\mu} (Ad_X \mu)_a^b W_{-\mu} W_{\mu} \delta_b W_{-\mu} =
(Ad_{\lambda_{\mu}X})_a^b (Ad_X (-\mu))_b^c \delta_c,
\eez
so that
\beq
(Ad_{\lambda_{\mu t}  Y}  \mu t)_a^b = ((Ad_Y (-\mu t))^{-1})_a^b \sim
\delta_a^b + t\mu^d U^b_{ad}(Y).
\l{3.14a}
\eeq
Furthermore,
\bey
(\delta_a V_{\mu t})[Y] \sim -it \mu^b \delta_a H_b(Y);\\
-i V_{\mu}[Y] H_a(Y) \sim -iH_a(Y) - t \mu^b H_b(Y) H_a(Y),
\eey
while the right-hand side of eq.\r{3.14} reads
\bez
<f, [-iH_a(Y) - it\mu^d U^b_{ad}(Y) H_b(Y) - it\mu^c\delta_c H_c(Y)  -
t\mu^c H_a(Y) H_c(Y)] g>.
\eez
We see  that  the property  \r{3.14}  implies the following algebraic
relation:
\beq
<f, ([H_a(Y), H_b(Y)] - i\delta_aH_b(Y) + i\delta_b H_a(Y)) g>
= iU^c_{ab}(Y) <f, H_c(Y) g>
\l{3.15}
\eeq
or
\beq
<f, [H_a(Y) - i\delta_a; H_b(Y) - i\delta_b] g>
= iU^c_{ab}(Y) <f, (H_c(Y) - i\delta_c) g>
\l{3.16}
\eeq
for $f,g \in {\cal F}_Y^0$.

{\bf 5.} Let us show that the algebraic property \r{3.16} implies  the
group property \r{3.12}.  First of all, let us obtain eq.\r{3.14} from
eq.\r{3.16}.

{\bf Proposition 3.1.} {\it Let property \r{3.16} be  satisfied.  Then
eq.\r{3.14} is also satisfied.}

{\bf Proof.} One should check that
\beb
<f, (-i   (V^0_{\mu   t}[Y])^+  (Ad_{\lambda_{\mu  t}  Y}  \mu  t)_a^b
H_b(\lambda_{\mu t} Y) V^0_{\mu t}[Y] -
(V^0_{\mu t}[Y])^+ (\delta_a V^0_{\mu t})[Y]) g >
= - i <f, H_a(Y) g>
\l{3.17}
\eeb
for all $f,g \in {\cal F}_Y^0$.

For $t=0$,  eq.\r{3.17}  is satisfied.  Consider the $t$-derivative of
the left-hand side of eq.\r{3.17}. First, eq.\r{3.12a} implies that
\bez
i \frac{\partial}{\partial t} V^0_{\mu t} [Y] =
\mu^a H_a(\lambda_{\mu t} Y) V^0_{\mu t} [Y],
\eez
while the operator $(\delta_a V^0_{\mu t})[Y]$ satisfies the following
equation:
\beb
\nu^a i \frac{\partial}{\partial t} (\delta_a V^0_{\mu t})[Y] =
\mu^b \frac{\partial}{\partial\alpha}|_{\alpha=0}  H_b(\lambda_{\mu t}
\lambda_{\nu\alpha} Y) V^0_{\mu t}[Y] + \mu^b H_b(\lambda_{\mu t}Y)
\nu^a (\delta_a V^0_{\mu t})[Y]
\l{3.18}
\eeb
Therefore,
\beq
\frac{d}{dt} [(V_{\mu t}^0[Y])^+ (\delta_a V^0_{\mu t})[Y]]
= -i (V_{\mu t}^0[Y])^+ (\delta_a W_{\mu t} H_b)(Y) V^0_{\mu t}[Y].
\eeq
Furthermore, eq.\r{3.14a} implies the following relation:
\bez
\frac{d}{dt} Ad_{\lambda_{\mu  t}  Y} \mu t = \frac{d}{dt}
(Ad_Y (-\mu t))^{-1}
= -
(Ad_Y (-\mu t))^{-1}
\frac{d}{dt}
(Ad_Y (-\mu t))
(Ad_Y (-\mu t))^{-1}
\eez
Eq. \r{3.8} implies that
\bez
\frac{d}{dt} (Ad_{\lambda_{\mu t} Y} \mu t)^a_b =
- (Ad_{\lambda_{\mu t} Y} \mu t)^a_c \mu^d U^b_{cd}  (\lambda_{\mu  t}
Y).
\eez
Combining all the terms,  we find that the derivative of the left-hand
side of eq.\r{3.17} is
\bey
<f, (V^0_{\mu t}[Y])^+ \{
- [(Ad_{\lambda_{\mu t} Y} \mu t)_a^b H_b(\lambda_{\mu t} Y);
\mu^c H_c(\lambda_{\mu t}Y)]
- i\mu^c  \delta_c  H_b(\lambda_{\mu  t} Y) (Ad_{\lambda_{\mu t}Y} \mu
t)_a^b +
\\
i(Ad_{\lambda_{\mu    t}    Y}    \mu     t)_a^c     \mu^d
U^b_{cd}(\lambda_{\mu t} Y) H_b(\lambda_{\mu t}Y)
+ i (W_{-\mu t} \delta_a W_{\mu t} H_b) (\lambda_{\mu t} Y) \}
(V^0_{\mu t} [Y]) g>
\eey
However, it  follows  from  eq.\r{3.16} that this expression vanishes.
Proposition is justified.

{\bf Proposition 3.2.} {\it The following property is satisfied:
\bez
(\delta_a V^0_{\mu t}) [Y] =  -  i  \int_0^t  d\tau  V^0_{(t-\tau)\mu}
(\lambda_{\mu t} Y \gets \lambda_{\mu\tau}Y) \mu^b
(\delta_a W_{\mu t} H_b)(Y) V^0_{\tau\mu} (\lambda_{\mu\tau}  Y  \gets
Y).
\eez
}

This property is a direct corollary of eq.\r{3.18}.

Let $\mu(\alpha)$  be  a smooth curve.  Note that for arbitrary function
$F(X)$ the  derivative  $\frac{d}{d\alpha}  F(\lambda_{\mu(\alpha)}X)$
can be presented as a linear combination of operators $\delta_a$,
\beq
\frac{d}{d\alpha} F(\lambda_{\mu(\alpha)}X)     =     \rho^a(\alpha,X)
\delta_a F(\lambda_{\mu(\alpha)}X).
\l{3.19}
\eeq

{\bf Proposition  3.3.}  {\it  For  all  $f  \in  {\cal  F}^0_{\lambda
\mu(\alpha)}X$, $g\in {\cal F}_X^0$
\beq
<f, i\frac{d}{d\alpha} V^0_{\mu(\alpha)}[X] g> =
<f, \rho^a(\alpha,X) H_a(\lambda_{\mu(\alpha)}X)  V^0_{\mu(\alpha)}[X]
g>
\l{3.20}
\eeq
}

{\bf Proof.} Consider the operator
\bez
(V^0_{t\mu(\alpha)} [X])^{-1} \frac{d}{d\alpha} V^0_{t\mu(\alpha)} [X]
\eez
Its time derivative has the form
\bez
(V^0_{t\mu(\alpha)} [X])^{-1} [\frac{d}{d\alpha};
\mu^a(\alpha) H_a(\lambda_{t\mu(\alpha)}X)]
V^0_{t\mu(\alpha)}[X]
\eez
Therefore,
\bez
(V^0_{t\mu(\alpha)} [X])^{-1} \frac{d}{d\alpha} V^0_{t\mu(\alpha)} [X]
=
\int_0^t d\gamma
V^0_{-\gamma \mu(\alpha)}(X \gets
\lambda_{\gamma \mu(\alpha)}X)
[\frac{d}{d\alpha}; \mu^a(\alpha) H_a(\lambda_{t\mu(\alpha)}X)]
V^0_{\gamma \mu(\alpha)}(\lambda_{\gamma \mu(\alpha)}X \gets X).
\eez
Thus, eq.\r{3.20} is equivalent to
\beb
<f,
\rho^a(\alpha,X) H_a(\lambda_{\mu(\alpha)}X) g>
= <f,
\int_0^t d\gamma
V^0_{(1-\gamma) \mu(\alpha)}(\lambda_{\mu(\alpha)} X \gets
\lambda_{\gamma \mu(\alpha)}X)
\frac{d}{d\alpha} [\mu^a(\alpha) H_a(\lambda_{t\mu(\alpha)}X)]
\\ \times
V^0_{(\gamma-1) \mu(\alpha)}(\lambda_{\gamma    \mu(\alpha)}X    \gets
\lambda_{\mu(\alpha)} X) g>.
\l{3.21}
\eeb
for all $f,g \in {\cal F}_X^0$.

Making use   of   the  justified  relation  \r{3.14},  one  takes  the
right-hand side of eq.\r{3.21} to the form
\beb
\int_0^1 d\gamma \frac{d\mu^a}{d\alpha}  (Ad_{\lambda_{\mu}X}  (\gamma
\mu))_a^b H_b(\lambda_{\mu} X)
- i              \int_0^1              d\gamma               (\delta_a
V^0_{(1-\gamma)\mu}[\lambda_{\gamma\mu}X]          V^0_{(\gamma-1)\mu}
(\lambda_{\gamma\mu} X \gets X) \frac{d\mu^a}{d\alpha} \\
+ \int_0^1    d\gamma    V^0_{(1-\gamma)\mu}   (\lambda_{\mu}X   \gets
\lambda_{\gamma\mu}X) \mu^a                         [\frac{d}{d\alpha}
H_a(\lambda_{\gamma\mu}X)]
V^0_{(\gamma-1) \mu} (\lambda_{\gamma\mu} X \gets \lambda_{\mu} X).
\l{3.22}
\eeb
The second term can be rewritten with the help of proposition 3.2 as
\bez
- \int_0^1 d\overline{\gamma} \int_0^{1-\overline{\gamma}} d\tau
V^0_{(1-\overline{\gamma}-\tau)\mu} [\lambda_{(\tau+ \overline{\gamma}\mu})
X] \mu^b (\delta_a W_{\tau\mu} H_b)(\lambda_{\overline{\gamma} \mu} X)
V^0_{(\tau + \overline{\gamma} -1) \mu} [\lambda_{\mu} X]
\frac{d\mu^a}{d\alpha},
\eez
where substitution  $\gamma  \rightarrow  \overline{\gamma}$  has been
made. Let us perform a shift of integration variable $\tau$,  $\tau  =
\gamma-\overline{\gamma}$, so that the integral will be transformed as
\bez
- \int_0^1 d\gamma \int_0^{\gamma} d\overline{\gamma}
V^0_{(1-\gamma) \mu} [\lambda_{\gamma\mu} X] \mu^b
(\delta_a W_{(\gamma - \overline{\gamma})\mu} H_b)(\lambda_{\gamma\mu}
X) V^0_{(\gamma-1) \mu} [\lambda_{\mu} X] \frac{d\mu^a}{d\alpha}.
\eez
We see that the sum of the second and the third term will vanish if
\beq
\int_0^{\gamma} d{\overline{\gamma}}          \mu^b          (\delta_a
W_{\overline{\gamma} \mu}            H_b)(\lambda_{(\gamma           -
\overline{\gamma})\mu}
X) \frac{d\mu^a}{d\alpha}       =       \mu^b        \frac{d}{d\alpha}
H_b(\lambda_{\gamma \mu(\alpha)} X).
\l{3.23}
\eeq
Here the  substitution  $\overline{\gamma}  \leftrightarrow  \gamma  -
\overline{\gamma}$ is  performed.  The  first term of the sum \r{3.22}
coincides with the left-hand side of eq.\r{3.21} if
\beq
\rho^b(\alpha,X) =     \int_0^1     d\gamma     \frac{d\mu^a}{d\alpha}
(Ad_{\lambda_{\mu} X} (\gamma \mu))_a^b.
\l{3.24}
\eeq
Definition \r{3.19} of $\rho^b$
implies that relation \r{3.24} is equivalent to
\beq
\frac{d}{d\alpha} f(\lambda_{\mu(\alpha)}X)         =         \int_0^1
d\overline{\gamma} \frac{d\mu^a}{d\alpha}         (Ad_{\lambda_{\mu}X}
(\overline{\gamma} \mu))_a^b (\delta_b f)(\lambda_{\mu(\alpha)}X)
\l{3.25}
\eeq
for arbitrary function $f$.  Eq.\r{3.23} is a corollary  of  the  more
general statement
\beq
\frac{d}{d\alpha} f(\lambda_{\gamma \mu(\alpha)}X) =
\int_0^{\gamma} (\delta_a      W_{\overline{\gamma}      \mu}       f)
(\lambda_{(\gamma - \overline{\gamma}) \mu} X) \frac{d\mu^a}{d\alpha}
\l{3.26}
\eeq
provided that $\mu^b H_b = f$. Since
\bez
(\delta_a W_{\overline{\gamma} \mu} f)(\lambda_{ (\gamma -
\overline{\gamma})\mu}X) =
(W_{-\overline{\gamma}\mu} \delta_a           W_{\overline{\gamma}\mu}
f)(\lambda_{\mu}X) =   (Ad_{\lambda_{\mu}X}  \overline{\gamma}\mu)_a^b
(\delta_b f)(\lambda_{\mu}X),
\eez
eq.\r{3.25} is a corollary of eq.\r{3.26}. To check relation \r{3.26},
note that
\bez
\frac{d}{d\gamma} W_{-\gamma  \mu}  \frac{d}{d\alpha} W_{\gamma \mu} =
W_{-\gamma\mu} \frac{d\mu^a}{d\alpha} \delta_a W_{\gamma\mu}.
\eez
Therefore,
\bez
W_{-\gamma  \mu}  \frac{d}{d\alpha} W_{\gamma \mu}
= \int_0^{\gamma}  d\overline{\gamma}
W_{-\overline{\gamma}\mu} \frac{d\mu^a}{d\alpha}
\delta_a W_{\overline{\gamma}\mu}
\eez
and
\bez
\frac{d}{d\alpha} (W_{\gamma \mu} f)(X) =
\int_0^{\gamma}  d\overline{\gamma}
W_{(\gamma -\overline{\gamma}) \mu} \frac{d\mu^a}{d\alpha}
\delta_a W_{\overline{\gamma}\mu}.
\eez
We obtain eq.\r{3.26}. Proposition is proved.

{\bf Proposition 3.4.} {\it Property \r{3.12} is satisfied.}

{\bf Proof.} Let $\mu(\tau)$ be such a function that
\beq
\lambda_{\mu(\tau)} X = \lambda_{\mu_1\tau} \lambda_{\mu_2} X.
\l{3.27}
\eeq
Let us show that
\beq
<f, (V^0_{\mu_1\tau}
(\lambda_{\mu_1\tau} \lambda_{\mu_2} X \gets
\lambda_{\mu_2}X))^+
V^0(\lambda_{\mu(\tau)} X \gets X) g>
= <f, V^0_{\mu_2} (\lambda_{\mu_2} X \gets X) g>.
\l{3.28}
\eeq
For $\tau  =  0$,  property  \r{3.28}  is  obviously  satisfied.   The
$\tau$-derivative of   the   left-hand   side   vanishes   because  of
proposition 3.3 and property $\rho^a = \mu_1^a$. Proposition is proved.

{\bf 6.}
Thus, we have understood that property \r{3.16}  is  a  necessary  and
sufficient condition for satisfying the group relation.  Let us check
eq.\r{3.16} for Weyl quantization.

Relation \r{3.16} can be rewritten as
\beq
<f,
[i\delta_{\Lambda_a} - \frac{1}{2} \Xi^2 \Lambda_a - \Lambda_a^{(1)};
i\delta_{\Lambda_b} - \frac{1}{2} \Xi^2 \Lambda_b - \Lambda_b^{(1)}]
g> = - i U^c_{ab} <f,
(i\delta_{\Lambda_c} - \frac{1}{2} \Xi^2 \Lambda_c - \Lambda_c^{(1)})
g>.
\l{3.29}
\eeq
It follows from the results of Appendix A that
\bez
[i\delta_{\Lambda_a} - \frac{1}{2} \Xi^2 \Lambda_a;
i\delta_{\Lambda_b} - \frac{1}{2} \Xi^2 \Lambda_b]
= i
i\delta_{\{ \Lambda_a; \Lambda_b\} } - \frac{1}{2} \Xi^2
(U^c_{ab} \Lambda_c))
\eez
Since $<f, \Xi \Lambda_c \tilde{g} > = 0$,
while $\delta_A B = \{A, B\}$, making use of property \r{3.5a},
one takes property \r{3.29} to the following form
\beq
\frac{i}{2} [  \Xi  U^c_{ab};  \Xi  \Lambda_c]  -  i  \{  \Lambda_a  ;
\Lambda_b^{(1)} \} - i \{ \Lambda_a^{(1)}; \Lambda_b\} =
i U^c_{ab} \Lambda_c^{(1)}.
\l{3.30}
\eeq
It follows from eq.\r{2.a} that
eq.\r{3.30} is satisfied if and only if
\beq
\{ \Lambda_a; Re \Lambda_b^{(1)} \} + \{ Re \Lambda_a^{(1)}; \Lambda_b
\} = U^c_{ab} Re \Lambda_c^{(1)};
\l{3.31}
\eeq
\beq
\frac{1}{2} \{  U^c_{ab};  \Lambda_c  \}  +  \frac{1}{2} \{ \lambda_a;
U^e_{eb} \} +  \frac{1}{2}  \{  U^d_{da};  \Lambda_b\}  =  \frac{1}{2}
U^c_{ab} U^d_{dc}.
\l{3.32}
\eeq
Eq.\r{3.31} is a restriction on the real part of quantum correction to
constraint. We  see  that  if  the  Weyl  quantization  is  used,  the
leading-order semiclassical purposes no  quantum  corrections  to  the
real part are necessary: the case $Re \Lambda_e^{(1)} = 0$ is possible.

Let us show that eq.\r{3.32} is automatically satisfied.  Consider the
Jacobi identity
\bez
[[\delta_a; \delta_b]; \delta_c] + [ [ \delta_b; \delta_c]; \delta_a] +
[ [ \delta_c; \delta_a]; \delta_b] = 0.
\eez
Making use of eq.\r{3.5a}, we find that
\bez
[U^d_{ab} \delta_d; \delta_c]
+ [U^d_{bc} \delta_d; \delta_a] + [U^d_{ca} \delta_d; \delta_b] = 0.
\eez
Applying eq.\r{3.5a} once again,  one takes the Jacobi identity to the
form
\bez
\{ \Lambda_c; U^d_{ab} \} \delta_d
+ \{ \Lambda_a; U^d_{bc} \} \delta_d +
\{ \Lambda_b; U^d_{ca} \} \delta_d +
U^d_{ab} U^e_{dc} \delta_e +
U^d_{bc} U^e_{da} \delta_e +
U^d_{ca} U^e_{db} \delta_e = 0.
\eez
Since vector fields $\delta_c$ are  linearly  independent  (the  gauge
group is assumed to act free on the phase space), one has
\bez
\{ \Lambda_c; U^e_{ab} \}
+ \{ \Lambda_a; U^e_{bc} \}  +
\{ \Lambda_b; U^e_{ca} \}  +
U^d_{ab} U^e_{dc}  +
U^d_{bc} U^e_{da}  +
U^d_{ca} U^e_{db}  = 0.
\eez
Consider the partial trace; let $c=e$ and sum over $e$. We obtain that
two last terms cancel each other and
\bez
\{ \Lambda_e; U^e_{ab} \} + \{ \Lambda_a; U^e_{be} \}
+ \{ \Lambda_b; U^e_{ea} \} + U^d_{ab} U^e_{de} = 0.
\eez
This relation coincides with \r{3.32}.

Thus, the algebraic property \r{3.29} is checked.  We see that if  the
Weyl quantization  is  used,  there  are  no  quantum anomalies in the
leading order of semiclassical theory.

On the other hand,  it is known from QFT that quantum anomalies  arise
in the one-loop approximation. This exactly corresponds to the leading
order of semiclassical approximation.

A possible source of  QFT  anomalies  may  be  as  follows.  The  Weyl
quantization cannot  be  applied to QFT systems:  one usually use Wick
ordering. There are also divergent  counterterms  to  the  Lagrangian.
This implies  that the quantum correction $\Lambda_e^{(1)}$ appears to
be not  only  nonzero  but  also  divergent.  Eq.\r{3.31}  is  then  a
nontrivial relation providing the cancellation of quantum anomalies.

\section{Semiclassical observables}

{\bf 1.}  In  classical mechanics of constrained systems,  observables
are such functions $O$ on classical phase space that commute with  all
the constraints on the constraint surface
\beq
\{ O, \Lambda_a \} = 0, \qquad {\rm provided \qquad that} \qquad
\Lambda_b = 0.
\l{4.1}
\eeq
Let us apply the corresponding quantum observable of the form
\bez
h\hat{O} = O(\sqrt{h} \hat{q}, \sqrt{h} \hat{p})
\eez
to the wave packet \r{2.12}. The result $\tilde{\Phi} = h\hat{O} \Phi$
will be of the analogous form
\bez
\tilde{\Phi}(q) = c e^{\frac{i}{h}S} e^{\frac{i}{h}P(q\sqrt{h} - Q)}
\tilde{f} (q-Q/\sqrt{h})
\eez
with
\bez
\tilde{f}(\xi) =       O(Q+\sqrt{h}\xi,       P       -      i\sqrt{h}
\frac{\partial}{\partial \xi}) f(\xi).
\eez
We see that a semiclassical wave function $(S,P,Q;f)$ is taken to
$(S,P,Q; \tilde{f})$ with the same $(S,P,Q)$.  In the leading order in
$h$ the function $f$ is multiplied by the classical  value  $O(Q,P)$ of
the observable. In the next order, one has
\bez
\tilde{f} = Of + \sqrt{h} \Xi O f + O(h),
\eez
see Appendix  A.  We  see  that  the  operator  $\Xi  O$  is the first
nontrivial contribution to the classical observable  $O$  and  may  be
viewed as a semiclassical observable.

Note that  zero-norm  semiclassical wave functions are indeed taken to
zero-norm states.  Namely,  let $<f,f>=0$.  Then for all $g$ from  the
domain of $(\Xi O)^+$
\bez
<g, \Xi O f> = (g, \prod_{a=1}^M (2\pi \delta(\Xi \Lambda_a)) \Xi O f)
= ((\Xi O)^+ g, \prod_{a=1}^M (2\pi \delta(\Xi \Lambda_a)) f) = 0,
\eez
since
\bez
[\Xi \Lambda_a; \Xi O ] = -i \{ \Lambda_a; O \}.
\eez
Therefore, the operator $\Xi O$ can be reduced to  factorspace  ${\cal
F}_X =  {\cal  F}_X^0/\sim$.  Any  bounded  function  of this operator
$\varphi(\Xi O)$ can be extended to $\overline{\cal F}_X$.

{\bf 2.}  It happens that it is sufficient to specify an observable on
the constraint surface $\Lambda_b = 0$ only in order  to  specify  the
operator $\Xi O$ in ${\cal F}_X$.

{\bf Proposition    4.1.}   {\it   Let   $O(Q,P)=0$,   provided   that
$\Lambda_b(Q,P) = 0$, $b=\overline{1,M}$. Then $\Xi O f$ has zero norm.
}

{\bf Proof.}  It  follows  from  the condition of proposition that the
linear form
\beq
\frac{\partial O}{\partial P_s} \delta P_s +
\frac{\partial O}{\partial Q_s} \delta Q_s
\l{4.1c}
\eeq
vanishes provided that
\beq
\frac{\partial\Lambda_b}{\partial P_s} \delta P_s +
\frac{\partial \Lambda_b}{\partial Q_s} \delta Q_s = 0.
\l{4.1b}
\eeq
Formula \r{4.1b}    specifies    $2n-M$-dimensional    subspace     of
$2n$-dimensional space $\{(\delta Q, \delta P)\}$, since the operators
$\Xi \Lambda_b$ are linearly independent because of free action of the
gauge group.  Choose such a basis $\{ (\delta Q^{(i)},  \delta P^{(i)}
), i = \overline{1,2n} \}$ that vectors $\{ (\delta Q^{\alpha}, \delta
P^{\alpha}), \alpha   =   \overline{M+1,2n}   \}$   satisfy  condition
\r{4.1b}. The  first  $M$  vectors   $(\delta   Q^{(a)},   \delta
P^{(a)})$, $a=\overline{1,M}$  should be chosen in such a way that
\bez
\frac{\partial\Lambda_b}{\partial P_s} \delta P_s
+ \frac{\partial\Lambda_b}{\partial Q_s} \delta Q_s = \delta_d^a
\l{4.1d}
\eez
Expand $(\delta Q, \delta P)$ as a linear combination
\bez
(\delta Q,  \delta  P)  =  \sum_{i=1}^n  b_i  (\delta Q^{(i)},  \delta
P^{(i)}).
\eez
The linear form \r{4.1c} can be presented as
\beq
\sum_{i=1}^n A_ib_i.
\l{4.1e}
\eeq
Let $b_{\alpha} =1$ for some $M+1 \le \alpha 2n$  and  other  $b_i=0$.
Then expression \r{4.1e} should vanish,  so  that  $A_{\alpha}  =  0$,
$\alpha = \overline{M+1,2n}$. One has
\bez
\frac{\partial O}{\partial P_s} \delta P_s +
\frac{\partial O}{\partial Q_s} \delta Q_s =
\sum_{a=1}^M A_a b_a =
\sum_a A_a
(\frac{\partial \Lambda_a}{\partial P_s} \delta P_s +
\frac{\partial \Lambda_a}{\partial Q_s} \delta Q_s).
\eez
This means that
\beq
\Xi O = \sum_{a=1}^M A_a \Xi \Lambda_a.
\l{4.1d1}
\eeq
However, the  states  $\Xi\Lambda_a f$ have zero norm.  Proposition is
proved.

{\bf 3.} Let us show that gauge-equivalent  semiclassical  states  are
taken by the operator $O + \sqrt{h} \Xi O$ to gauge-equivalent. Let
\bez
(S^0,P^0,Q^0; f^0) \sim (S^{\tau},P^{\tau},Q^{\tau}: f^{\tau}),
\eez
$(P^{\tau},Q^{\tau})$ satisfy  eq.\r{2.17},  $S^{\tau}$ be of the form
\r{2.16}, $f^{\tau}$ be a solution of eq.\r{2.17m}.  First, one should
check that
\beq
O(Q^{\tau},P^{\tau}) = O(Q^{0},P^{0})
\l{4.2}
\eeq
or $\frac{d}{d\tau} O(Q^{\tau},P^{\tau}) = 0$.  However,  the property
is equivalent to $\{O, \Lambda_a\} = 0$ and therefore satisfied.

Let us check that states
\bez
(S^0,P^0,Q^0; \Xi O f^0) \sim (S^{\tau},P^{\tau},Q^{\tau}: \Xi O
f^{\tau})
\eez
are gauge-equivalent. Consider the gauge transformation
$(S^{\tau},P^{\tau},Q^{\tau}; g^{\tau})$ of the state
$(S^0,P^0,Q^0; \Xi O f^0)$.  $g^{\tau}$ is  than  a  solution  of  the
Cauchy problem  for eq.\r{2.17m} with the initial condition $g^0 = \Xi
O f^0$.  The difference $\zeta^{\tau} = g^{\tau}  -  \Xi  O  f^{\tau}$
satisfies the following equation
\bez
(i \frac{d}{d\tau} - \mu^a
[\frac{1}{2} \Xi^2 \Lambda_a + \Lambda_a^{(1)}])
\zeta^{\tau} =
- [i\frac{d}{d\tau}  -  \frac{1}{2}  \mu^a  \Xi^2\Lambda_a;   \Xi   O]
f^{\tau}.
\eez
It follows from Appendix A that
\bez
[i\frac{d}{d\tau} - \frac{1}{2} \mu^a \Xi^2\Lambda_a;  \Xi O] = \Xi \{
\mu^a \Lambda_a; O \}.
\eez
Since $\{\Lambda_a; O\} = 0$ on the constraint surface,
\bez
\Xi \{ \mu^a \Lambda_a ; O \} = A^a \Xi \Lambda_a
\eez
for some   coefficients   $A^a$.   Therefore,   the   wave    function
$\zeta^{\tau}$ satisfies the following equation
\bez
(i \frac{d}{d\tau} - \mu^a
[\frac{1}{2} \Xi^2 \Lambda_a + \Lambda_a^{(1)}]
\zeta^{\tau} =
\Xi \Lambda_a \chi^{\tau}_a
\eez
for some  $\chi_a^{\tau}$.  For  the  inner  product
$<\zeta^{\tau}, \zeta^{\tau}>$, analogously to eq.\r{3.5} one has
\beq
i \frac{d}{d\tau} <\zeta^{\tau}, \zeta^{\tau}>
= <\zeta^{\tau}, \Xi \Lambda_a \zeta^{\tau}>
- <\Xi\Lambda_a \zeta^{\tau}, \zeta^{\tau}> = 0.
\l{4.2d}
\eeq
Therefore, $\zeta^{\tau}$ is a state of a zero norm.  This means  that
wave functions  $g^{\tau}$ and $\Xi O f^{\tau}$ are equivalent.  Thus,
equivalent semiclassical states are indeed taken to equivalent by  the
semiclassical observable $\Xi O$.

{\bf 4.}  It  happens  also  that  semiclassical  observables  $\Xi O$
possesses also the following geometric interpretation  (cf.\c{Sh1,Sh2}).
By $K_{S,P,Q}$  we  denote the operator taking the function $f$ to the
wave function $\Phi$ \r{2.12},
\beq
(K^h_{S,P,Q} f)(q) = \frac{1}{h^{M/4}} e^{\frac{i}{h}S} e^{\frac{i}{h}
P(q\sqrt{h} - Q)} f(q-Q/\sqrt{h}).
\l{4.3}
\eeq
Consider the shift of classical variables of the  order  $O(h^{1/2})$,
$S \to S+ \sqrt{h} \delta S$, $P \to P + \sqrt{h}\delta P$, $Q \to Q +
\sqrt{h} \delta Q$.  The operator $K_{S,P,Q}$ will transform  then  as
follows,
\bez
(K^h_{S+\sqrt{h} \delta S, P+ \sqrt{h} \delta P, Q+ \sqrt{h} \delta Q}
f = const K^h_{S,P,Q} e^{i\Omega[\delta P,\delta Q]} f,
\eez
where $\Omega[\delta P,\delta Q]$ is the following linear  combination
of coordinate and momenta operators,
\bez
(\Omega[\delta P,  \delta  Q]  f)(\xi)  =  [\delta  P  \xi  - \delta Q
\frac{1}{i} \frac{\partial}{\partial \xi}] f(\xi).
\l{4.4}
\eez
$\Omega$ is  an  operator-valued  differential form:  a tangent vector
$(\delta P,\delta Q)$ to the phase space is mapped to an operator; the
mapping is  linear.  The  form  $\Omega$  is  an  important  geometric
characteristics of the operator $K^h_{S,P,Q}$.

We see that linear combinations of coordinate and  momentum  operators
can be  expressed  via  the  operators $\Omega$.  However,  $\delta P,
\delta Q$  should  obey  additional  restrictions  since  the  mapping
$K^h_{S,P,Q}$ is  defined  only  if  $(P,Q)$ belongs to the constraint
surface
\bez
\Lambda_a(Q,P) = 0.
\eez
This means that
\beq
\frac{\partial\Lambda_a}{\partial Q} \delta Q
+ \frac{\partial\Lambda_a}{\partial P} \delta P = 0.
\l{4.5}
\eeq
Under condition \r{4.5},  the  operator  $\Omega[\delta  P,\delta  Q]$
transforms zero-norm states to zero-norm states, since
\bez
[\Omega(\delta P,\delta Q); \Xi \Lambda_a] =
i (\frac{\partial\Lambda_a}{\partial Q} \delta Q
+ \frac{\partial\Lambda_a}{\partial P} \delta P) = 0.
\eez
Let $O$  be  a  classical  observable.  We  see that the corresponding
semiclassical observable $\Xi O$ can be presented as
\bez
\Xi O  =  \Omega[\frac{\partial  O}{\partial  Q};   -   \frac{\partial
O}{\partial P}].
\eez
Note that the tangent vectors
\bez
\delta_O P = - \frac{\partial O}{\partial Q}; \qquad
\delta_O Q = \frac{\partial O}{\partial P}.
\eez
corresponds to the Hamiltonian vector field $\delta_O$  on  the  phase
space which   is  generated  by  the  classical  observable  $O$.  One
therefore has
\beq
\Xi O = - \Omega[\delta_O (P,Q)].
\l{4.6}
\eeq

\section{Semiclassical transformations}

Quantum observables $\hat{O}$ can be  also  viewed  as  generators  of
one-parametric transformation    groups   $e^{-i\hat{O}t}$.   Let   us
investigate their  analogs  in  the   semiclassical   mechanics.   Let
$\hat{O}$ depend on the small parameter $h$ as
\beq
\hat{O} = \frac{1}{h} O(\sqrt{h} \hat{q}, \sqrt{h} \hat{p}) +
 O_1 (\sqrt{h} \hat{q}, \sqrt{h} \hat{p}) + ...
\l{5.i}
\eeq
It happens that unitary condition for  the  constrained  systems  make
necessary adding  a  quantum  correction  $O_1$  with  a  nontrivial
imaginary part.

Let us apply the operator $e^{-i\hat{O}t}$ to the  semiclassical  wave
function $\Phi$ \r{2.12}. Consider the wave function
\bez
\Phi^t = e^{-i\hat{O}t} \Phi
\eez
obeying the Cauchy problem
\beq
i \dot{\Phi}^t = \hat{O} \Phi^t, \qquad \Phi^0 = \Phi.
\l{5.0}
\eeq
Analogously to section 2, substitution
\bez
\Phi^t (q) = const e^{\frac{i}{h}S^t} e^{\frac{i}{h} P^t (q\sqrt{h}  -
Q^t)} f^t(q - Q^t/\sqrt{h})
\eez
gives us in the leading order in $h$ the following system of equations,
\beq
\dot{S}^t = P^t \dot{Q}^t - O(Q^t,P^t); \qquad
\dot{Q}^t = \frac{\partial O}{\partial P}(Q^t,P^t); \qquad
\dot{P}^t = - \frac{\partial O}{\partial Q}(Q^t,P^t).
\l{5.1}
\eeq
\beq
i \dot{f}^t = \left[
\frac{1}{2} (\Xi^2 O)(Q^t,P^t) + O_1 (Q^t,P^t)
\right] f^t
\l{5.2}
\eeq
Note that the classical trajectory $Q^t,P^t$ lies  on  the  constraint
surface, provided  that  $\{  O,  \Lambda_a  \}  = 0$ on this surface.
Therefore, one can define the transformation $u_t:  {\cal X} \to {\cal
X}$ taking  the  initial  data  $(S^0,Q^0,P^0)$ for eqs.\r{5.1} to the
solution $(S^t,Q^t,P^t)$.  By $U_t^0(u_t X \gets X):  {\cal F}_X^0 \to
{\cal F}_{u_tX}^0$  we  denote  the  operator  taking the initial wave
function $f^0$ to the solution $f^t$. Let us investigate the unitarity
property of the infinitesimal operator.

\subsection{Unitarity problem}

One can  check  that  the operator $U^0_t(u_tX \gets X)$ conserves the
norm analogously to section 3. Let us investigate the commutator
\bez
[ i \frac{d}{dt} - \frac{1}{2} (\Xi^2 O)(u_tX) ;
(\Xi \Lambda_a)(u_tX)]
\eez
which has the form
\bez
i \Xi \{O; \Lambda_a\}
\eez
according to   appendix   A.  However,  the  Poisson  bracket  $\{  O;
\Lambda_a\}$ vanishes   on   the   constraint   surface.    Therefore,
proposition 4.1 implies that
\bez
\Xi \{O; \Lambda_a\} = A_a^b \Xi \Lambda_b,
\eez
where coefficient  functions  $A_a^b(X)$ are uniquely defined from the
relation
\beq
\delta_{\{O;\Lambda_a\}} = A_a^b \delta_{\Lambda_b}.
\l{5.1c}
\eeq
Therefore,
\bez
[i \frac{d}{dt} - \frac{1}{2} (\Xi^2O)(u_tX); (\Xi \Lambda_a)(u_tX)] =
i A_a^b(u_tX) (\Xi \Lambda_b)(u_tX)
\eez
and
\beq
[i\frac{d}{dt} - \frac{1}{2} (\Xi^2 O)(u_tX);
e^{i\rho^a (\Xi \Lambda_a)(u_tX)}]
= i A_a^b(u_tX) \rho^a \frac{\partial}{\partial \rho^b}
e^{i\rho^c (\Xi \Lambda_c)(u_tX)}.
\l{5.1d}
\eeq
Let $f_1^t$,  $f_2^t$  be  solutions  of  eq.\r{5.0}.  Analogously  to
eq.\r{3.3}, we find
\bez
i \frac{\partial}{\partial t} <f_1^t, f_2^t> =
[O_1(u_tX) - O_1^*(u_tX)] <f_1^t,  f_2^t> -  i  A_a^a  (u_tX)  <f_1^t,
f_2^t>.
\eez
Thus, a   semiclassical   transformations   conserves   the   norm  of
semiclassical state  if  and  only  if  $O_1$  contains  a  nontrivial
imaginary part
\beq
Im O_1(X) = \frac{1}{2} A_a^a (X).
\l{5.3}
\eeq
Under this condition zero-norm states are taken to  zero-norm  states,
so that  the  operators  $U_t^0(u_t^0X  \gets  X)$  can  be reduced to
factorspace. Namely,  introduce the unitary operators $U_t(u_tX  \gets
X): {\cal F}_X \to {\cal F}_{u_tX}$ by the following definition
\beq
U_t(u_tX \gets  X)  [f^0]  = [U_t^0(u_tX\gets X) f^0],  \qquad f^0 \in
{\cal F}_X^0
\l{5.4}
\eeq
which is correct.  Operators \r{5.4} can be extended to the completion
of ${\cal F}_X$,
\bez
\overline{U}_t(u_tX \gets X):  \overline{\cal F}_X \to  \overline{\cal
F}_{u_tX}.
\eez
We see  that  a  semiclassical  transformation  may  be  viewed  as an
automorphism of the semiclassical bundle.

\subsection{Gauge invariance}

The purpose of this subsection is to show that gauge-equivalent  states
are taken to gauge-equivalent.  It is convenient to introduce a notion
of a {\it pre-semiclassical} bundle with  base  $\cal  X$  and  fibres
${\cal F}_X^0$ and consider the sections of this bundle. Remind that a
section of a bundle is specified if for each $X\in {\cal  X}$  a  wave
function $\chi_X^0  \in {\cal F}_X^0$ is chosen;  certain requirements
on smoothness of dependence of $\chi_X^0$ on $X$ may be imposed.

{\it
A section   $\chi^0$   of   the  pre-semiclassical  bundle  is  called
gauge-invariant if for all $\mu$, $X\in {\cal X}$ the property
\beq
\chi^0_{\lambda_{\mu}X} \sim  V_{\mu}^0   (\lambda_{\mu}   X\gets   X)
\chi_X^0
\l{5.5}
\eeq
is satisfied.
}

{\bf Proposition 5.1.} {\it Property \r{5.5} is equivalent to}
\beq
<f^0, [i\delta_a - H_a(X)]\chi_X^0> = 0, \qquad
{\rm for all} \qquad f^0 \in {\cal F}_X^0.
\l{5.6}
\eeq

{\bf Proof.} Property \r{5.5} can be rewritten as
\beq
<f^0; \chi^0_X   -   V^0_{\mu   t}(X    \gets    \lambda_{-\mu    t}X)
\chi_{\lambda_{-\mu t} X}^0>  = 0.
\eeq
Consider the limit $t\to 0$. One has
\bey
V^0_{\mu t} ( X\gets \lambda_{-\mu t} X) \sim
1 - i \mu_a t H_a(X) + o(t), \\
\chi^0_{\lambda_{-\mu t} X} \sim \chi_X^0 - t\mu_a \delta_a \chi^0_X +
o(t).
\eey
We obtain relation \r{5.6}.

Let us check the implication \r{5.6} $\to$ \r{5.5}. The wave function
\bez
\zeta_t = \chi^0_{\lambda_{\mu t} X} -
V^0_{\mu t} (\lambda_{\mu t} X \gets X) \chi_X^0
\eez
satisfies the equation
\bez
i \dot{\zeta}_t  = H_a(\lambda_{\mu t} X) \zeta_t + [i\mu_a \delta_a -
H_a(\lambda_{\mu t} X) ] \chi^0_{\lambda_{\mu t} X},
\eez
so that
\bez
i \frac{d}{dt} <\zeta_t, \zeta_t> = 0
\eez
analogously to \r{4.2d}. Proposition is proved.

{\bf Proposition 5.2.} {\it  Let $f_1 \in \overline{\cal F}_{X_1}$,
$f_2 \in \overline{\cal F}_{X_2}$,  $f_1 \ne 0$,  $f_2\ne 0$. Then the
semiclassical states $(X_1,f_1)$ and $(X_2,f_2)$ are  gauge-equivalent
if and  only  if  for  all  gauge-invariant  sections  $\chi^0$ of the
pre-semiclassical bundle the relation
\beq
< [\chi_{X_1}^0]; f_1> =
< [\chi_{X_2}^0]; f_2>
\l{5.7}
\eeq
is satisfied.
}

{\bf Proof.}   Let   $(X_1,f_1)   \sim  (X_2,f_2)$,  $\chi_X^0$  be  a
gauge-invariant section. Then
\bez
X_2 = \lambda_{\mu} X_1, \qquad
f_2 = V_{\mu}(X_2 \gets X_1) f_1, \qquad
[\chi_{X_2}^0] = [V_{\mu}^0(X_2\gets X_1) \chi_{X_1}^0]
\eez
for some $\mu$. One should check that
\bez
<[\chi_{X_1}^0]; f_1> =
<[V_{\mu}^0 (X_2\gets X_1) \chi_{X_1}^0];
V_{\mu}(X_2\gets X_1) f_1>.
\eez
This is true because of unitarity of $V_{\mu}(X_2\gets X_1)$.

Let $(X_1,f_1) \not\sim (X_2,f_2)$. One should consider two cases.

1. $X_1\not\sim X_2$.

Then $X_1,X_2$ belong to different gauge  orbits.  One  therefore  can
choose such  an  invariant section $\chi^0$ that $\chi_{X_1}^0 = F^0$,
$\chi_{X_2}^0 =  0$,  where  $f^0$  is  an  arbitrary  wave  function.
Therefore,
\bez
< [f^0]; f_1> = 0
\eez
for all $f^0$, so that $f_1= 0$.

2. $X_1 \sim X_2$,  so that $X_2 = \lambda_{|mu} X_1$, but $f_2 \ne
V_{\mu}(X_2\gets X_1) f_1$.  Choose such an invariant section $\chi^0$
that $\chi^0_{X_2} = f^0$ be an arbitrary nonzero wave function.
One has
\bez
[\chi^0_{X_2}] = [V_{\mu}^0(X_2\gets X_1) \chi^0_{X_1}].
\eez
Therefore,
\bez
<[\chi^0_{X_1}]; f_1> =
<[\chi^0_{X_2}]; V_{\mu}(X_2\gets X_1)f_1>.
\eez
For some $f^0$, one has
\bez
<[f^0]; V_{\mu}(X_2\gets X_1) f_1> \ne <[f^0]; f_1>,
\eez
so that
\bez
<[\chi^0_{X_1}];f_1>
\ne
<[\chi^0_{X_2}];f_2>.
\eez
Proposition is proved.

Call the section $\chi^t$ of the form
\beq
\chi^t_{u_tX} = U_t^0(u_tX \gets X) \chi^0_X
\l{5.7b}
\eeq
as a semiclassical transformation of the section $\chi^0$.

{\bf Proposition 5.3.} {\it  Semiclassical  transformation  takes  any
gauge-equivalent section to a gauge-invariant section if and only if
\beq
(X_1,f_1) \sim (X_2,f_2) \Rightarrow
(u_tX_1; U_t(u_t X_1 \gets X_1) f_1) \sim
(u_tX_2; U_t(u_t X_2 \gets X_2) f_2).
\l{5.7a}
\eeq
}

{\bf Proof.}   Let   the   semiclassical   transformation   take   any
gauge-invariant section  to  a  gauge-invariant  and  $(X_1,f_1)  \sim
(X_2,f_2)$. Let us show
\beq
(u_tX_1; U_t(u_tX_1 \gets X_1) f_1)
\sim (u_tX_2; U_t(u_tX_2 \gets X_2) f_2)
\l{5.8a}
\eeq
According to proposition 5.3, one should check that
\beq
<[\chi^0_{u_tX_1}]; U_t(u_tX_1 \gets X_1) f_1)> =
<[\chi^0_{u_tX_2}]; U_t(u_tX_2 \gets X_2) f_2)>
\l{5.8}
\eeq
for all  gauge-invariant  sections  $\chi^0$.  Property   \r{5.8}   is
equivalent to
\bez
<[U^0_{-t}(X_1 \gets u_tX_1) \chi^0_{u_tX_1}]; f_1> =
<[U^0_{-t}(X_2 \gets u_tX_2) \chi^0_{u_tX_2}]; f_2>
\eez
or
\beq
<[\chi_{X_1}^{-t}]; f_1> =
<[\chi_{X_2}^{-t}]; f_2>.
\l{5.9}
\eeq
However, the  section  $[\chi_X^{-t}]$   is   gauge-invariant,   while
$(X_1;f_1) \sim (X_2;f_2)$.  Thus,  eq.\r{5.9} is satisfied.  Property
\r{5.8a} is checked.

Let us  suppose  now  that  implication  \r{5.7a}  takes  place.   Let
$\chi^0_X$ be   a  gauge-invariant  section.  Show  that  the  section
$\chi^t_X$ is gauge-invariant, i.e.
\bez
(X, [\chi^t_X]) \sim (\lambda_{\mu}X, [\chi^t_{\lambda_{\mu}X}])
\eez
or
\beq
(X, [U_t^0(X \gets u_{-t}X) \chi^0_{u_{-t}X}])
\sim
(\lambda_{\mu} X, [U_t^0( \lambda_{\mu}
X \gets u_{-t} \lambda_{\mu} X) \chi^0_{u_{-t} \lambda_{\mu} X}])
\l{5.10}
\eeq
However,
\bez
(u_{-t}X, [\chi^0_{u_{-t}X}])
\sim
(u_{-t}\lambda_{\mu} X, [\chi^0_{u_{-t} \lambda_{\mu}X}])
\eez
since $u_{-t}X \sim u_{-t} \lambda_{\mu} X$, while section $\chi^0$ is
gauge-invariant, so   that   implication  \r{5.7a}  implies  \r{5.10}.
Proposition is proved.

To check that  equivalent  states  are  taken  to  equivalent  by  the
semiclassical transformation,   it   is   sufficient   to   show  that
gauge-invariant sections are taken to gauge-invariant, i.e.
\beq
<f^0, [i\delta_a - H_a(X)] \chi_X^t> = 0
\qquad {\rm for \qquad all} \qquad f^0.
\l{5.11}
\eeq
under condition \r{5.6}. Denote
\beq
\delta_O f(X) = \frac{d}{dt}|_{t=0} f(u_tX).
\eeq

{\bf Proposition 5.4.} {\it \r{5.6} $\Rightarrow$ \r{5.11} if and only
if
\beq
<f, [i\delta_a -  H_a(X);  i\delta_O  -  \frac{1}{2}  (\Xi^2  O)(X)  -
O_1(X)] g> = -i <f, A_a^b(X) (i\delta_b-H_b(X)) g>.
\l{5.12}
\eeq
where $A_a^b(X)$ are defined from eq.\r{5.1c}.
}

{\bf Proof.}  Let  property  \r{5.11}  be satisfied for all $\chi_X^0$
obeying eq.\r{5.6}.  Let us obtain eq.\r{5.12}.  Relation \r{5.11} can
be rewritten as
\beq
<f^0, (i\delta_a - H_a(X)) U_t^0(X \gets u_{-t}X) \chi^0_{u_{-t}X} = 0.
\l{5.13}
\eeq
Consider the limit $t\to 0$. The leading order in $t$ gives us trivial
result \r{5.6}. The next order leads us to the following relations,
\bey
\chi^0_{u_{-t}X} \sim \chi^0_X  - t \delta_O\chi_X^0 + o(t);
\\
U^0_t(X \gets u_{-t}X) \sim 1 - it [\frac{1}{2} (\Xi^2 O)(X) + O_1(X)]
+ o(t),
\eey
so that
\bez
<f^0, (i\delta_a - H_a(X)) (i\delta_O -  \frac{1}{2}  (\Xi^2  O)(X)  -
O_1(X)) \chi_X^0> = 0.
\eez
On the other hand, for
\bez
g_X^0 = (i\delta_a - H_a(X)) \chi_X^0
\eez
one has
\bey
<f^0, (i\delta_O - \frac{1}{2} (\Xi^2 O)(X) - O_1(X)) g_X^0> =
\\
(f^0, [ \prod_c (2\pi \delta(\Xi\Lambda_c(X)));
i\delta_O - \frac{1}{2} (\Xi^2 O)(X) - O_1(X)]
g_X^0) = \\
i (f^0, \prod_c (2\pi \delta(\Xi\Lambda_c(X))) i A_a^a(X) g_X^0) = 0,
\eey
here properties \r{5.6}, \r{5.1d} are used. Therefore,
\beq
<f^0, [i\delta_a - H_a(X);
i\delta_O - \frac{1}{2} (\Xi^2 O)(X) - O_1(X)] \chi_X^0> = 0.
\l{5.14}
\eeq
The commutator entering to this expression has the structure
\beq
-\delta_{\{\Lambda_a; O\} } + B_a(X)
\l{5.15}
\eeq
for some  operator  function  $B_a(X)$.  It follows from \r{5.1c} that
eq.\r{5.14} takes the form
\bez
<f^0, (A_a^b(X) \delta_b + B_a(X)) \chi_X^0> = 0.
\eez
Eq.\r{5.6} implies that
\bez
A_a^b(X) iH_b(X) = B_a(X)
\eez
in the weak sense. Thus, eq.\r{5.15} implies \r{5.12}.

Let eq.\r{5.12} be satisfied.  Check the property  \r{5.11}.  Consider
the wave function
\beq
\zeta^a[t;X] = [U_t(u_tX \gets X)]^{-1} \phi^a[t,u_tX].
\l{5.15a}
\eeq
Let us obtain an equation for $\zeta^a$. Notice that the wave function
$\chi^t_{u_tX}$ satisfies the equation
\beq
i \frac{d}{dt}   \chi^t_{u_tX}   =   [\frac{1}{2}  (\Xi^2  O)(u_tX)  +
O_1(u_tX)] \chi^t_{u_tX}
\l{5.16}
\eeq
being a   corollary  of  relation  \r{5.7b}.  The  left-hand  side  of
eq.\r{5.16} can be presented as
\bez
(i \frac{\partial}{\partial t} + i \delta_O) \chi^t_Y|_{Y=u_tX},
\eez
so that
\bez
i \frac{\partial}{\partial t} \chi_Y^t =  [\frac{1}{2}  (\Xi^2O)(Y)  +
O_1(Y) - i\delta_O] \chi_Y^t.
\eez
Therefore, the function $\phi^a[t,u_tX]$ obeys the equation
\bez
i \frac{d}{dt} \phi^a[t,u_tX] =
i\delta_O \phi^a[t,u_tX] +
(i\delta_a - H_a(Y))
(\frac{1}{2} (\Xi^2O)(Y)    +    O_1(Y)    -     i\delta_O))|_{Y=u_tX}
\chi^t_{u_tX}.
\eez
It follows from eq.\r{5.12} that it can be rewritten as
\bez
i \frac{d}{dt} \phi^a[t,u_tX] = O_2(u_tX) \phi^a[t,u_tX] +
i A_a^b(u_tX) \phi^b[t,u_tX].
\eez
Thus,
\beq
i \frac{d}{dt}  \zeta^a[t,u_tX]   =   [U_t(u_tX   \gets   X)]^{-1}   i
A_a^b(u_tX) [U_t(u_tX \gets X)] \zeta^b[t,X].
\l{5.17}
\eeq
The operator entering to the right-hand side is bounded, so that there
exists a unique solution for the Cauchy problem for eq.\r{5.17}, which
can be presented as a strongly convergent series,
\bez
\zeta^a[t,X] = (T\exp \{ - \int_0^t  d\tau
[U_{\tau}(u_{\tau}X  \gets X)]^{-1} A(u_{\tau}X)
[U_{\tau}(u_{\tau}X  \gets X)]\} \zeta)^a[0,X].
\eez
Each term of the series has zero norm,  so that  $<\zeta^a,\zeta^a>  =
0$. Proposition 5.4 is proved.

\subsection{Check of infinitesimal properties}

Let us  check  property  \r{5.12}  analogously  to subsection 3.3.  It
follows from the results of appendix A that
\beq
[i\delta_a - \frac{1}{2} \Xi^2 \Lambda_a - \Lambda_a^{(1)};
i\delta_O - \frac{1}{2} \Xi^2O - O_1] =
i
(i\delta_{\{\Lambda_a; O\}}  -  \frac{1}{2}  \Xi^2\{\Lambda_a;  O\}  +
\delta_O \Lambda_a^{(1)} - \delta_a O_1)
\l{5.18}
\eeq
Let us   first   calculate   the   operator   $\frac{1}{2}   \Xi^2\{O;
\Lambda_a\}$. It follows from \r{5.1c} that
\bez
\frac{\partial \{O;\Lambda_a\}}{\partial Q_j} = A_a^b
\frac{\partial \Lambda_b}{\partial Q_j};
\qquad
\frac{\partial \{O;\Lambda_a\}}{\partial P_j} = A_a^b
\frac{\partial \Lambda_b}{\partial P_j};
\eez
on the constraint  surface.  According  to  proposition  4.1  (formula
\r{4.1d}), one can write:
\bey
\frac{\partial^2 \{O,\Lambda_a\} }{\partial Q_i \partial Q_j}
=
\frac{\partial A_a^b}{\partial Q_i}
\frac{\partial \Lambda_b}{\partial Q_j}
+
A_a^b \frac{\partial^2 \Lambda_b}{\partial Q_i \partial Q_j}
+ \lambda_Q^{acj} \frac{\partial \Lambda_c}{\partial Q_i};\\
\frac{\partial^2 \{O,\Lambda_a\} }{\partial P_i \partial Q_j}
= \frac{\partial A_a^b}{\partial P_i}
\frac{\partial \Lambda_b}{\partial Q_j}
+ A_a^b \frac{\partial^2 \Lambda_b}{\partial P_i \partial Q_j}
+ \lambda_Q^{acj} \frac{\partial \Lambda_c}{\partial P_i}\\
\frac{\partial^2 \{O,\Lambda_a\} }{\partial Q_i \partial P_j}
= \frac{\partial A_a^b}{\partial Q_i}
\frac{\partial \Lambda_b}{\partial P_j}
+
A_a^b \frac{\partial^2 \Lambda_b}{\partial Q_i \partial P_j}
+ \lambda_P^{acj} \frac{\partial \Lambda_c}{\partial Q_i}\\
\frac{\partial^2 \{O,\Lambda_a\} }{\partial P_i \partial P_j}
= \frac{\partial A_a^b}{\partial P_i}
\frac{\partial \Lambda_b}{\partial P_j}
+
A_a^b \frac{\partial^2 \Lambda_b}{\partial P_i \partial P_j}
+ \lambda_P^{acj} \frac{\partial \Lambda_c}{\partial P_i}
\eey
for some functions $\lambda_Q^{ac}$, $\lambda_P^{ac}$.
Thus,
\bez
\Xi^2\{O; \Lambda_a\} = A_a^b \Xi^2\Lambda_b + \Xi A_a^b \Xi \Lambda_b
+ \Xi \Lambda_c (\lambda_Q^{acj} \xi_j +  \lambda_P^{acj}  \frac{1}{i}
\frac{\partial}{\partial\xi_j}.
\eez
However, the matrix elements of the operator $\Xi \Lambda_c \hat{B}_c$
are zero for arbitrary  $\hat{B}_c$.  Thus,  the  operator  $\Xi^2\{O,
\Lambda_a\}$ is equivalent in sense of matrix elements to
\bez
\Xi^2 \{O;  \Lambda_a\}  \sim  A_c^b \Xi^2\Lambda_b + [\Xi A_a^b;  \Xi
\Lambda_b].
\eez
Expression \r{5.18} can be transformed  as
\bez
-iA_a^b (i\delta_b - \frac{1}{2} \Xi^2 \Lambda_b) +  \frac{i}{2}  [\Xi
A_a^b; \Xi \Lambda_b] + i\delta_O\Lambda_a^{(1)} - i\delta_a O_1.
\eez
Eq.\r{5.12} is then satisfied if and only if
\beq
iA_a^b \Lambda_b^{(1)} = \frac{1}{2} \{ A_a^b; \Lambda_b\}
+ i \{O; \Lambda_a^{(1)} \} - i \{ \Lambda_a; O_1\}.
\l{5.19a}
\eeq
Decompose this relation into real and imaginary parts:
\beq
iA_a^b Re  \Lambda_b^{(1)}  =  i  \{  O;  Re  \Lambda_a^{(1)}  -  i \{
\Lambda_a; Re O_1\};
\l{5.19}
\eeq
\beq
- \frac{1}{2} A_a^b U^c_{cb} = \frac{1}{2} \{ A_a^b;  \Lambda_b  \}  -
\frac{1}{2} \{ O; U^c_{ca} \} + \frac{1}{2} \{\Lambda_a; A_c^c\} = 0.
\l{5.20}
\eeq
It happens that eq.\r{5.20} is automatically satisfied.

Namely, the Jacobi identity
\bez
[[\delta_a,\delta_b],\delta_O] + [[\delta_b,\delta_O],\delta_a]
+ [[\delta_O,\delta_a],\delta_b] = 0.
\eez
It can be rewritten as
\bez
U^c_{ab} A_c^d  \delta_d + \delta_O U^d_{ab} \delta_d + A_b^c U^d_{ca}
\delta_d + \delta_aA_b^d \delta_d - A_a^c U^d_{cb} \delta_d - \delta_b
A_a^d \delta_d = 0.
\eez
Since $\delta_d$ are independent operators, one has
\bez
U^c_{ab} A_c^d + A^c_b U^d_{ca} - A^c_a U^d_{cb} +
\{ O; U^d_{ab} \} + \{ \Lambda_a; A_b^d\} - \{ \Lambda_b; A_a^d\} = 0.
\eez
Multiplying this relation  by  the  $\delta$-symbol  $\delta_b^d$,  we
obtain eq.\r{5.20}.

Thus, gauge-equivalent states are indeed taken to gauge-equivalent via
the semiclassical transformation generated by the observable $O$ under
condition \r{5.19}.  If  $Re \Lambda_a^{(1)} = 0$,  one can choose $Re
O_1 = 0$.  Otherwise,  the non-anomaly condition for $O_1$ \r{5.19} is
obtained.

\subsection{Relationship with BRST-BFV approach}

Let us  compare  the  obtained  results  \r{5.3},  \r{5.19a}  with the
definition of the observable in the  BRST-BFV  quantization.  In  this
approach, a  B-extension of an observable $\hat{O}$ is considered.  It
is looked for in the following form
\bez
\hat{O}_B =   \hat{O}   +   ...   +   \hat{O}^{nb_1...b_n}_{a_1...a_n}
\overline{\Pi}_{b_1} ... \overline{\Pi}_{b_n}
\frac{\partial}{\partial \overline{\Pi}_{a_1}} ...
\frac{\partial}{\partial \overline{\Pi}_{a_n}} + ...
\eez
in such a way that
\bez
[\hat{O}_B; \hat{\Omega}_0] = 0; \qquad
\hat{O}_B^+ = \hat{O}_B.
\eez
The purpose of  this  subsection  is  to  show  that  if  the  quantum
observable $\hat{O}$  possesses a B-extension,  then relations \r{5.3}
and \r{5.19a} will be  satisfied.  We  suppose  that  the  B-extension
depends on  the  small parameter $h$ semiclassically (eq.\r{2.b}),  so
that the coefficient operators $\hat{O}^n$ depend on $h$ as
\bez
\hat{O}^{nb_1...b_n}_{a_1...a_n} =
h^{n-1}  O^{nb_1...b_n}_{a_1...a_n}
(\sqrt{h} \hat{q}, \sqrt{h} \hat{p})
h^{n}  O^{n b_1...b_n}_{1 a_1...a_n}
(\sqrt{h} \hat{q}, \sqrt{h} \hat{p}) + ...
\eez
Let us investigate first the condition $[\hat{O}_B;  \hat{\Omega}_0] =
0$. One has
\bez
\hat{O}_B \hat{\Omega}_0        =       \hat{O}       \hat{\Omega}_a^1
\frac{\partial}{\partial \overline{\Pi}_a} + ...; \qquad
\hat{\Omega}_0 \hat{O}_B = \hat{\Omega}_a^1 \hat{O}
\frac{\partial}{\partial \overline{\Pi}_a} +
\hat{\Omega}_a^1 \hat{O}^{1a}_{a_1}
\frac{\partial}{\partial \overline{\Pi}_{a_1}} + ...
\eez
where ... are terms containing $\overline{\Pi}_b$. The $\overline{\Pi}$
- $\frac{\partial}{\partial \overline{\Pi}}$ ordering is chosen. Thus,
one has the following condition
\bez
[\hat{O}; \hat{\Lambda}_a] = \hat{\Lambda}_b \hat{O}_a^{1b}.
\eez
In the leading order of the  semiclassical  approximation  results  of
appendix A imply that
\beq
- i \{O; \Lambda_a \} = \Lambda_b O^{1b}_a;
\l{5.23}
\eeq
\beq
-i (\{O_1;\Lambda_a\} + \{O; \Lambda_a^{(1)} \})
= \Lambda_b^{(1)} O^{1b}_a + \Lambda_b O^{1b}_a -
\frac{i}{2} \{ \Lambda_b; O^{1b}_a \}.
\l{5.24}
\eeq
Let us consider  the  condition  $\hat{O}_B^+  =  \hat{O}_B$.  In  the
leading orders in $h$, this implies that
\beq
O^* = O; \qquad O^{1b*}_a = - O^{1b}_a; \qquad
O_1^* - O_1 + O_a^{1a*} = 0.
\l{5.25}
\eeq
Comparison of  formulas  \r{5.23}  and \r{5.1c} gives us the following
relation
\bez
O^{1b}_a = - iA_a^b.
\eez
which is valid on the constrained surface.
Thus,
\bez
Im O_1 = \frac{1}{2} A_a^a.
\eez
On the  constraint  surface  eq.\r{5.24}  coincides with eq.\r{5.19a}.
Thus, relations \r{5.3} and \r{5.19a} are checked.

\subsection{Equivalent observables}

The constructed  semiclassical  transformation  depends  not  only  on
values of  the  classical observable $O$ on the constraint surface but
also on the off-constraint-surface values.  Namely, if one adds to the
observable $O$ terms $\alpha^a\Lambda_a$, even the classical equations
\r{5.1} will change. It happens, however, that classical states $u_tX$
corresponding to transformations generated by observables $O$ and $O +
\alpha^a\Lambda_a$ are gauge-equivalent. The purpose of this subsection
is to  show  that  an  analogous  statement  is  valid  also  for  the
semiclassical transformation.

Let $\hat{C}$ and $\hat{B}$ be two semiclassical observables,
\bey
\hat{C} = \frac{1}{h} C(\sqrt{h} \hat{q}, \sqrt{h}\hat{p})
+  C_1(\sqrt{h} \hat{q}, \sqrt{h}\hat{p}) + ...; \\
\hat{B} = \frac{1}{h} B(\sqrt{h} \hat{q}, \sqrt{h}\hat{p})
+  B_1(\sqrt{h} \hat{q}, \sqrt{h}\hat{p}) + ...; \\
\eey
such that $C=B$ and $Re C_1 = Re B_1$ on the constraint surface. Let
\bey
(X,f^0) \mapsto (u_t[B]X, U_t^0(u_t[B] X \gets X; B) f^0);\\
(X,f^0) \mapsto (u_t[C]X, U_t^0(u_t[B] X \gets X; C) f^0)
\eey
be semiclassical transformations generated by the observables $B$  and
$C$.

{\bf Proposition 5.5}
{\it The following relation is satisfied:
\beq
(u_t[B]X, [U_t^0(u_t[B] X \gets X; B) f^0])
\sim (u_t[C]X, [U_t^0(u_t[C] X \gets X; C) f^0]).
\l{5.26}
\eeq
}

To prove property \r{5.26},  it is sufficient according to proposition
5.2 to  check  that  for  all  gauge-invariant sections $\chi_X^0$ the
property
\beq
<\chi^0_{u_t[B]X}, U_t^0(u_t[B] X \gets X; B) f^0>
= <\chi^0_{u_t[C]X}, U_t^0(u_t[C] X \gets X; C) f^0>
\l{5.27}
\eeq
is satisfied. Relation \r{5.27} can be rewritten as
\beq
<f^0; U_{-t}^0(X \gets u_t[B] X; B) \chi^0_{u_t[B]X}>
= <f^0; U_{-t}^0(X \gets u_t[C] X; C) \chi^0_{u_t[C]X}>
\l{5.28}
\eeq
Denote
\bez
\chi_X^{-t}[B] = U^0_{-t} (X \gets u_t[B]X; B) \chi^0_{u_t[B]X},
\qquad
\chi_X^{-t}[C] = U^0_{-t} (X \gets u_t[C]X; C) \chi^0_{u_t[C]X}.
\eez
One should check that there is an equivalence,
\beq
\chi_X^{-t}[B] \sim \chi_X^{-t}[C].
\l{5.28a}
\eeq
Note that the considered sections obey the following equations,
\bey
\frac{\partial}{\partial t} \chi_X^{-t}[B] =
i (\frac{1}{2} (\Xi^2 B)(X) + B_1(X) - i\delta_B) \chi_X^{-t}[B];\\
\frac{\partial}{\partial t} \chi_X^{-t}[C] =
i (\frac{1}{2} (\Xi^2 C)(X) + C_1(X) - i\delta_C) \chi_X^{-t}[C].
\l{5.29}
\eey
Therefore, the difference
\bez
\rho_X^{-t} = \chi_X^{-t} [B] - \chi_X^{-t}[C]
\eez
satisfies the following equation
\beq
\frac{\partial}{\partial t} \rho_X^{-t} =
i (\frac{1}{2} (\Xi^2  B)(X)  +  B_1(X)  -  i\delta_B)  \rho_X^{-t}  +
\gamma_X^{-t}
\l{5.30}
\eeq
with
\beq
\gamma_X^{-t} = i
(\frac{1}{2} (\Xi^2  O)(X)  +  O_1(X)  -  i\delta_O)  \chi_X^{-t}[C].,
\l{5.30a}
\eeq
where $O=B-C$,  so that $O=0$ on the constrained surface.  One  should
check that $\rho_X^{-t} \sim 0$ or
\beq
<\rho^{-t}_{u_{-t}[B]X};\rho^{-t}_{u_{-t}[B]X}> = 0.
\l{5.31}
\eeq
The time derivative of the left-hand side of eq.\r{5.31} is
\bez
i <\rho^{-t}_{u_{-t}[B]X};\gamma^{-t}_{u_{-t}[B]X}>
- i <\gamma^{-t}_{u_{-t}[B]X};\rho^{-t}_{u_{-t}[B]X}>.
\eez
Since at  the  initial  moment  of time relation \r{5.31} is obviously
satisfied, it is sufficient to check
\bez
\gamma_X^{-t} \sim 0.
\eez
First of all, note that eq.\r{4.1d} implies that
\bez
\delta_O = A_a \delta_a;
\eez
or
\bez
\frac{\partial O}{\partial     P_i}     =      A_a      \frac{\partial
\Lambda_a}{\partial P_i};
\qquad
\frac{\partial O}{\partial     Q_i}     =      A_a      \frac{\partial
\Lambda_a}{\partial Q_i}.
\eez
Therefore, analogously to subsection 5.3, we obtain
\bez
\Xi^2 O = A_a \Xi^2\Lambda_a + [\Xi A_a; \Xi\Lambda_a] + \Xi \Lambda_a
\hat{B}_a
\eez
for some operator $\hat{B}_a$. Therefore,
\bez
\gamma_X^{-t} \sim   i  (\frac{1}{2}  A_a(X)  (\Xi^2  \Lambda_a)(X)  -
\frac{i}{2} \{A_a;  \Lambda_a\}(X)  +  O_1(X)  -  i  A_a(X)  \delta_a)
\chi_X^{-t}[C].
\eez
Making use of eq.\r{5.11}, we find that
\bez
\gamma_X^{-t} \sim   i
(- \frac{i}{2} \{A_a;   \Lambda_a\}(X)    +    O_1(X)    -
i    A_a(X) \Lambda_a^{(1)}(X))
\chi_X^{-t}[C].
\eez
Thus, it is sufficient to show that the relation
\beq
\frac{1}{2} \{A_a;\Lambda_a\} + iO_1 + A_a \Lambda_a^{(1)} = 0
\l{5.32}
\eeq
is satisfied.  Let  us  decompose  condition  \r{5.32}  into  real and
imaginary parts. Let us find coefficients in eq.\r{5.1c}. One has
\bez
[\delta_O; \delta_{\Lambda_a}] = [A_c\delta_c;\delta_a] =
-A_c U^d_{ca} \delta_d - \{ \Lambda_a; A_c\} \delta_c,
\eez
so that
\bez
A_a^d = - A_c U^d_{ca} - \{A_a; \Lambda_d\}
\eez
and
\bez
Im O_1 = - \frac{1}{2} A_c U^d_{ca} - \frac{1}{2} \{ \Lambda_a;  A_a\}
= A_c \Lambda_c^{(1)} + \frac{1}{2} \{ A_a; \Lambda_a\}.
\eez
Eq.\r{5.32} takes the form $Re O_1 = 0$.  Therefore,  under  condition
$Re B_1  =  Re  C_1$  relation  \r{5.26}  is satisfied.  Semiclassical
transformations generated   by   observables   $B$   and    $C$    are
gauge-equivalent.

\subsection{Semiclassical transformations of semiclassical observables}

It happens  that  the  linear  combinations  of  operators  $\xi$  and
$\partial/\partial\xi$ transform under time evolution in a simple way.

Let $\delta P^t$,  $\delta  Q^t$  satisfy  the  variation  system  for
eqs.\r{5.1}:
\bey
\frac{d}{dt}\delta Q^t  =
\frac{\partial^2  O}{\partial P \partial P}
(Q^t,P^t) \delta P^t +  \frac{\partial^2  O}{\partial P \partial Q}
(Q^t,P^t) \delta Q^t; \\
\frac{d}{dt}\delta P^t  =
- \frac{\partial^2  O}{\partial Q \partial P}
(Q^t,P^t) \delta P^t - \frac{\partial^2  O}{\partial Q \partial Q}
(Q^t,P^t) \delta Q^t.
\eey

{\bf Proposition 5.6.} {\it  The following identity is satisfied:
\beq
\Omega[\delta P^t, \delta Q^t] U_t^0(u_tX \gets X)
= U^0_t(u_tX \gets X) \Omega[\delta Q^0,\delta P^0].
\l{5.33a}
\eeq
}

{\bf Proof.} Let $f^t = U_t^0(u_tX \gets X) f^0$, $g^0 = \Omega[\delta
Q^0,\delta P^0] f^0$, $g^t = U_t^0(u_tX \gets X) g^0$ is a solution of
the Cauchy problem for the equation
\beq
i \frac{\partial}{\partial t} g^t = [\frac{1}{2} (\Xi^2 O)(u_tX) +
O_1(u_tX)] g^t.
\l{5.33}
\eeq
One should  check that $g^t = \Omega[\delta q^t,\delta P^t]f^t$.  This
function indeed satisfies the initial condition, while
\bey
[i\frac{\partial}{\partial t} - \frac{1}{2} (\Xi^2O)(u_tX) - O_1(u_tX);
\Omega[\delta Q^t, \delta P^t]] = \\
i\Omega[
\frac{d}{dt}\delta Q^t  -
\frac{\partial^2  O}{\partial P \partial P}
(Q^t,P^t) \delta P^t -  \frac{\partial^2  O}{\partial P \partial Q}
(Q^t,P^t) \delta Q^t;
\frac{d}{dt}\delta P^t  +
\frac{\partial^2  O}{\partial Q \partial P}
(Q^t,P^t) \delta P^t + \frac{\partial^2  O}{\partial Q \partial Q}
(Q^t,P^t) \delta Q^t]
= 0,
\eey
so that  the  function  $\Omega[\delta Q^t,\delta P^t] f^t$ obeys also
eq.\r{5.33}. Proposition is proved.

Property \r{5.33a} which is satisfied not only for real  $\delta  Q^t,
\delta P^t$  but  also for the complex $\delta Q^t,\delta P^t$ is very
useful for constructing the quasi-Gaussian solutions of eq.\r{5.2}.

Namely, the Gaussian function
\bez
f^0(\xi) = const e^{\frac{i}{2} \xi_i \alpha_{ij} \xi_j}
\eez
can be geometrically interpreted in terms of the Maslov complex germ.
For the  quantum mechanics,  the corresponding theory was developed in
\c{Maslov2}. It  was  generalized  to  the  case   of   the   abstract
semiclassical mechanics in \c{Sh2}.

A Maslov  complex  germ is a $n$-dimensional plane in the complexified
$2n$-dimensional tangent space to the phase space,
\bez
{\cal G}_{\alpha} = \{ (\delta Q,\delta P) | \delta P_i =  \alpha_{ij}
\delta Q_j\}.
\eez
One has
\bez
(\delta Q,\delta  P)  \in  {\cal  G}_{\alpha}  \Leftrightarrow  \Omega
[\delta Q,\delta P] f^0 = 0.
\eez
Moreover, the property $\Omega[\delta X] f^0 = 0$ for  all  $\delta  X
\in {\cal  G}_{\alpha}$ uniquely specify the wave function $f^0$ up to
a multiplicative factor.

Let $u_{*t}[X]$ be a mapping of tangent spaces to the phase space.  It
follows from proposition 5.6 that
\bez
\Omega[u_{*t}[X] \delta  X]  f_t  =  0,  \qquad  \delta  X  \in  {\cal
G}_{\alpha}.
\eez
Thus, $f_t$ is a Gaussian function,  while  the  complex  germ  ${\cal
G}_{\alpha^t}$ is $u_{*t}[X]{\cal G}_{\alpha^0}$.

If the initial condition is a product of a polynomial by the exponent,
it can be presented as a sum of functions
\bez
\Omega [\delta X_1] ... \Omega [\delta X_n] f_0,
\eez
so that at time moment $t$ one obtains the function
\bez
\Omega[\delta X_1^t] ... \Omega[\delta X_n^t] f_t
\eez
with $\delta X_i^t = u_{*t}[X] \delta X_i$.

One can also develop the complex-germ theory for the function
$\prod_a (2\pi\delta(\Xi  \Lambda_a)) f^a(\xi)$ which is also Gaussian
(see \c{Sh2} for details).

\section{Composed semiclassical states}

We have  investigated  the  properties  of  the  wave  packet function
\r{2.12} corresponding to an "elementary" semiclassical state.  It  is
known from   quantum   mechanics   (see   appendix  B)  that  infinite
superpositions of states \r{2.12} may be also viewed as  semiclassical
solutions of  the semiclassical equations.  In particular,  WKB-states
can be obtained in such a way.

However, one should be careful:  sometimes the superposition  \r{B.11}
vanishes up  to $O(h^{\infty})$ and gives us therefore a trivial state
$\Psi = 0$.  There is also a gauge-like ambiguity  even  for  theories
without constraints: under certain transformations of $g$ the function
$\Psi$ does not vary.

\subsection{Superpositions of elementary semiclassical states}

Let $X(\alpha)=  (S(\alpha),   Q(\alpha),   P(\alpha))$,   $\alpha   =
(\alpha_1,...,\alpha_k)$ be  a $k$-dimensional surface embedded to the
base of the semiclassical bundle $\cal X$.  Let $g(\alpha,\xi)$  be  a
function of   the   class  ${\cal  S}({\bf  R}^{k+n})$.  Consider  the
superposition of the wave packets \r{2.12}
\beq
\Phi(q) = c \int d\alpha e^{\frac{i}{h}S(\alpha)}
e^{\frac{i}{h} P(\alpha) (q\sqrt{h} - Q(\alpha))}
g(\alpha, q - \frac{Q(\alpha)}{\sqrt{h}}).
\l{6.1}
\eeq

\subsubsection{Explicit form of the composed semiclassical state}

One can calculate  the  integral  \r{6.1}  explicitly  analogously  to
\c{MS}. First, notice that the integral \r{6.1} is exponentially small
if the distance between $q$ and the  surface  $Q(\alpha)$  is  of  the
order $O(h^{-1/2})$.  A  nontrivial  result  will  be  obtained  only if this
distance is   of   the   order    $O(h^{1/2})$,    i.e.    for    some
$\overline{\alpha}$ $q = h^{-1/2} Q(\overline{\alpha}) + \xi$,  $\xi =
O(1)$. Consider the substitution $\alpha = \overline{\alpha}  +  \beta
\sqrt{h}$. The integral will be taken to the form:
\beq
ch^{k/2} e^{\frac{i}{h} S(\overline{\alpha})}
e^{\frac{i}{h} P(\overline{\alpha}) (q\sqrt{h} - Q(\overline{\alpha})}
\int d\beta e^{\frac{i}{\sqrt{h}} \beta_j
( \frac{\partial S}{\partial \overline{\alpha}_j} -
P \frac{\partial Q}{\partial \overline{\alpha}_j} )}
e^{\frac{i}{2} \beta_i\beta_j (
\frac{\partial^2 S}
{\partial \overline{\alpha}_i \partial \overline{\alpha}_j}
- P \frac{\partial^2 Q}
{\partial \overline{\alpha}_i \partial \overline{\alpha}_j}
)}
e^{i \beta_j \frac{\partial P}{\partial \overline{\alpha}_j} \xi}
g(\alpha,\xi - \frac{\partial Q}{\partial \overline{\alpha}_j}\beta_j ),
\l{6.1m}
\eeq
here the higher-order terms is $h$ are omitted.
Notice that this integral contains a rapidly oscillating expression
$e^{\frac{i}{\sqrt{h}} \beta_j
( \frac{\partial S}{\partial \overline{\alpha}_j} -
P \frac{\partial Q}{\partial \overline{\alpha}_j} )}$.
Therefore, the integral is exponentially small, except for the case:
\beq
\frac{\partial S}{\partial \overline{\alpha}_j} =
P \frac{\partial Q}{\partial \overline{\alpha}_j}.
\l{6.6}
\eeq
This is  the  Maslov  isotropic  condition  \c{Maslov2}.  Under   this
requirement, one can simplify the integral \r{6.1m}:
\beq
c h^{k/2} e^{\frac{i}{h} S(\overline{\alpha})}
e^{\frac{i}{h} P(\overline{\alpha}) (q\sqrt{h} - Q(\overline{\alpha}))}
f(\alpha,q - \frac{Q(\overline{\alpha})}{\sqrt{h}});
\l{6.1xx}
\eeq
with
\bez
f(\alpha,\xi) = \int d\beta
e^{i \beta_j (\frac{\partial P}{\partial \overline{\alpha}_j} \xi
- \frac{\partial P}{\partial \overline{\alpha}_j}
\frac{1}{i} \frac{\partial}{\partial \xi})}
g(\alpha,\xi) =
\prod_{i=1}^k
\delta (\frac{\partial P}{\partial \overline{\alpha}_j} \xi
- \frac{\partial P}{\partial \overline{\alpha}_j}
\frac{1}{i} \frac{\partial}{\partial \xi})
g(\alpha,\xi).
\eez

\subsubsection{Semiclassical inner product}

Investigate the inner  product  $<\Phi,\Phi>$.  Let  us  make  use  of
formula \r{2.5}.  The  wave  function  \r{2.i1}  will  have  then  the
following form
\bez
\Phi^{\tau}(q) = c \int d\alpha e^{\frac{i}{h} S^{\tau}(\alpha)}
e^{\frac{i}{h} P^{\tau}(\alpha) (q\sqrt{h} - Q^{\tau}(\alpha))}
\chi^{\tau}(\alpha, q-     \frac{Q^{\tau}(\alpha)}{\sqrt{h}},     \Pi,
\overline{\Pi}).
\eez
The functions          $S^{\tau}(\alpha)$,         $P^{\tau}(\alpha)$,
$Q^{\tau}(\alpha)$ satisfy   system    \r{2.16},    \r{2.17},    while
$\chi^{\tau}$ obeys      eq.\r{2.i2}      and      $\chi^0(\alpha,\xi,
\Pi,\overline{\Pi}) =  g(\alpha,\xi)$.  For   $\tau   =   -1$   denote
$S^{\tau}(\alpha) \equiv S(\alpha,\mu)$,
$P^{\tau}(\alpha) \equiv P(\alpha,\mu)$,
$Q^{\tau}(\alpha) \equiv Q(\alpha,\mu)$,
$\chi^{\tau}(\alpha, \xi,      \Pi,       \overline{\Pi})       \equiv
\chi(\alpha,\mu,\xi,\Pi, \overline{\Pi})$.
Therefore, the wave function
\bez
\Psi(q) =   \int   \prod_{a=1}^M   d\mu_a   d\overline{\Pi}_a   d\Pi^a
e^{-\overline{\Pi}_a \Pi^a       +      i\mu_a      [\overline{\Pi}_a;
\hat{\Omega}_0]_+} \Phi(q).
\eez
entering to the inner  product  \r{2.5}  has  the  form  analogous  to
\r{6.1}:
\beq
\Psi(q) = c \int d\alpha d\mu e^{\frac{i}{h}S(\alpha,\mu)}
e^{\frac{i}{h} P(\alpha,\mu) (q\sqrt{h} - Q(\alpha,\mu))}
g(\alpha, \mu, q - \frac{Q(\alpha,\mu)}{\sqrt{h}}).
\l{6.2}
\eeq
with
\beq
g(\alpha,\mu,\xi) =  \int   \prod_{a=1}^M   d\overline{\Pi}_a   d\Pi^a
\chi(\alpha,\mu,\xi, \Pi,\overline{\Pi}).
\l{6.3}
\eeq
This is a Dirac wave  function  corresponding  to  the  state  \r{6.1}
\c{Sh}.

Let us      suppose      that      the     $k$-dimensional     surface
$(Q(\alpha),P(\alpha))$ contains  no  gauge-equivalent  states.   More
precisely, we         require         that        the        manifold
$(Q(\alpha,\mu),P(\alpha,\mu))$ is smooth,  $k+M$-dimensional, without
self-intersections. For such a case, consider the inner product
\bez
<\Phi,\Phi> \equiv (\Phi,\Psi)
\eez
which has the form
\beb
<\Phi,\Phi> =   |c|^2   \int   d\alpha   d\gamma  d\mu  e^{\frac{i}{h}
(S(\gamma,\mu) -    S(\alpha))}\\
\times \int    dq    g^*(\alpha,    q    -
\frac{Q(\alpha)}{\sqrt{h}}) e^{\frac{i}{h}P(\gamma,\mu)  (q\sqrt{h}  -
Q(\gamma,\mu)) - \frac{i}{h} P(\alpha) (q\sqrt{h} - Q(\alpha))}
g(\gamma,\mu, q- \frac{Q(\gamma,\mu)}{\sqrt{h}}).
\l{6.4}
\eeb
Analogously to subsection 2.2,  the integrand is exponentially  small,
except for  the  case  $|P(\alpha)  -  P(\gamma,\mu)|  = O(\sqrt{h})$,
$|Q(\alpha) -  Q(\gamma,\mu)|  =  O(\sqrt{h})$.  Therefore,  only  the
domain $|\gamma-  \alpha| = O(\sqrt{h})$,  $|\mu| = O(\sqrt{h})$ gives
rise to a nontrivial contribution to the integral \r{6.4}.  Thus,  one
should perform a substitution
\bez
\gamma = \alpha + \beta \sqrt{h},  \qquad \mu = \rho\sqrt{h}, \qquad q
= Q(\alpha) + \xi\sqrt{h}.
\eez
The inner product \r{6.4} takes the form
\beb
<\Phi,\Phi> = |c|^2 h^{\frac{M+n+k}{2}} \int d\alpha d\beta d\rho
\exp[\frac{i}{h}(S(\alpha+\sqrt{h}\beta, \rho\sqrt{h})  - S(\alpha)) -
P(\alpha + \sqrt{h}\beta, \rho\sqrt{h}) - Q(\alpha))]
\\ \times
\int d\xi g^*(\alpha,\xi)
e^{\frac{i}{\sqrt{h}} (P(\alpha+\sqrt{h}\beta,          \rho\sqrt{h})-
P(\alpha))\xi }   g(\alpha+   \beta   \sqrt{h},  \rho\sqrt{h},  \xi  +
\frac{Q(\alpha) - Q(\alpha+ \sqrt{h}\beta, \sqrt{h}\rho)}{\sqrt{h}})
\l{6.5}
\eeb
the singular term entering to the exponent is
\bez
\frac{i}{\sqrt{h}} (\frac{\partial   S}{\partial   \alpha}   \beta   +
\frac{\partial S}{\partial \mu} \rho -  P  \frac{\partial  Q}{\partial
\alpha} \beta - P \frac{\partial Q}{\partial \mu} \rho)
\eez
If it is nonzero,  the integral \r{6.5} contains a rapidly oscillating
factor and becomes therefore exponentially  small.  Thus,  one  should
impose the condition \r{6.6} and the following requirement:
\beq
\frac{\partial S(\alpha,0)}{\partial   \mu_a}   =   P   \frac{\partial
Q(\alpha,0)}{\partial \mu_a}.
\l{6.7}
\eeq
However, relations  \r{6.7}  is  automatically  satisfied  because  of
\r{2.16}, provided that
\beq
\Lambda_a (Q(\alpha),P(\alpha)) = 0.
\l{6.8}
\eeq
(cf. subsection 2.2).

If conditions  \r{6.6} and \r{6.7} are satisfied,  the integral \r{6.5}
is of the order $O(1)$, provided that the normalizing factor is chosen
to be
\bez
|c| = h^{-\frac{M+n+k}{4}}
\eez
Consider the limit $h\to 0$.
Notice that eq.\r{2.17} implies for $\mu_a=0$ that
\beq
\frac{\partial Q}{\partial\mu_a}       =       -        \frac{\partial
\Lambda_a}{\partial P};
\qquad
\frac{\partial P}{\partial\mu_a}       =            \frac{\partial
\Lambda_a}{\partial Q}.
\l{6.9}
\eeq
Differentiating eq.\r{2.16} with respect to $\tau$, one finds
\bez
\ddot{S}^0 - P^0\ddot{Q}^0 = \dot{P}^0\dot{Q}^0,
\eez
or
\bez
\frac{\partial^2S}{\partial\mu_a \partial\mu_b} -
P \frac{\partial^2 Q}{\partial\mu_a \partial\mu_b} =
\frac{\partial P}{\partial\mu_a}
\frac{\partial Q}{\partial\mu_b}
= \frac{\partial \Lambda_a}{\partial Q}
\frac{\partial \Lambda_b}{\partial P}
\eez
Differentiating eq.\r{6.7} with respect to $\alpha$, one finds:
\bez
\frac{\partial^2S}{\partial\mu_a \partial \alpha_i}
= \frac{\partial P}{\partial \alpha_i}
\frac{\partial Q}{\partial \mu_a} +
P \frac{\partial^2Q}{\partial\alpha_i \partial \mu_a}.
\eez
Furthermore, eq.\r{6.6} implies
\bez
\frac{\partial^2 S}{\partial\alpha_i \partial\alpha_j} =
\frac{\partial P}{\partial\alpha_i} \frac{\partial Q}{\partial\alpha_j}
+ P \frac{\partial^2 Q}{\partial\alpha_i \partial\alpha_j}.
\eez
We see that
\bez
\frac{\partial P}{\partial\alpha_j}
\frac{\partial Q}{\partial\alpha_i}
=
\frac{\partial Q}{\partial\alpha_j}
\frac{\partial P}{\partial\alpha_i}
\eez
since
$\frac{\partial^2S}{\partial\alpha_i \partial  \alpha_j}$  should   be
symmetric.

Thus, the exponent under conditions \r{6.6}, \r{6.7} takes the form
\bey
i [\frac{1}{2}
\beta_i
(\frac{\partial^2S}{\partial\alpha_i \partial\alpha_j}
- P \frac{\partial^2Q}{\partial\alpha_i \partial\alpha_j})
\beta_j
+
\beta_i
(\frac{\partial^2S}{\partial\alpha_i \partial\mu_a}
- P \frac{\partial^2Q}{\partial\alpha_i \partial\mu_a})
\rho_a
+
\frac{1}{2}
\rho_a
(\frac{\partial^2S}{\partial\mu_a \partial\mu_b}
- P \frac{\partial^2Q}{\partial\mu_a \partial\mu_b})
\rho_b
\\
-
(\beta_i \frac{\partial P}{\partial\alpha_i} +  \rho_a  \frac{\partial
P}{\partial\mu_a})
(\frac{\partial Q}{\partial\alpha_i}\beta_i +  \frac{\partial
Q}{\partial\mu_a} \rho_a) =
\\
i (\frac{1}{2} \beta_i \frac{\partial P}{\partial \alpha_i}
\frac{\partial Q}{\partial \alpha_j} \beta_j -
\beta_i \frac{\partial     P}{\partial     \alpha_i}    \frac{\partial
\Lambda_a}{\partial P}  \rho_a  +  \frac{1}{2}  \rho_a  \frac{\partial
\Lambda_a}{\partial Q} \frac{\partial \Lambda_b}{\partial P} \rho_b -
(\beta_i \frac{\partial P}{\partial\alpha_i} +  \rho_a
\frac{\partial \Lambda_a}{\partial Q})
(\frac{\partial Q}{\partial\alpha_i}\beta_i -  \frac{\partial
\Lambda_a}{\partial P} \rho_a) =\\
- \frac{i}{2} \beta_i \frac{\partial P}{\partial\alpha_i}
\frac{\partial Q}{\partial\alpha_j} \beta_j -
i \rho_a \frac{\partial\Lambda_a}{\partial       Q}       \frac{\partial
Q}{\alpha_i} \beta_i     +     \frac{i}{2}    \rho_a    \frac{\partial
\Lambda_a}{\partial Q} \frac{\partial \Lambda_b}{\partial P} \rho_b.
\eey
Under condition \r{6.8},  making use of eq.\r{6.9}, we obtain as $h\to
0$ that
\bey
<\Phi,\Phi>  \to \int d\alpha d\beta d\rho
e^{
- \frac{i}{2} \beta_i \frac{\partial P}{\partial\alpha_i}
\frac{\partial Q}{\partial\alpha_j} \beta_j -
i \rho_a \frac{\partial\Lambda_a}{\partial       Q}       \frac{\partial
Q}{\alpha_i} \beta_i     +     \frac{i}{2}    \rho_a    \frac{\partial
\Lambda_a}{\partial Q} \frac{\partial \Lambda_b}{\partial P} \rho_b
} \\
\times
(g, e^{
i (\frac{\partial         P}{\partial        \alpha_i}\beta_i        +
\frac{\partial\Lambda_a}{\partial Q} \rho_a) \xi}
e^{-
(\frac{\partial Q}{\partial\alpha_i} \beta_i -
\frac{\partial\Lambda_a}{\partial P}\rho_a)
\frac{\partial}{\partial\xi}
} g).
\eey
Making use of the Baker-Hausdorff formula and relation
\bez
\frac{\partial\Lambda_a}{\partial P}     \frac{\partial    P}{\partial
\alpha_j} +   \frac{\partial\Lambda_a}{\partial   Q}    \frac{\partial
Q}{\partial \alpha_j} = 0
\eez
which is  a  corollary  of  property  \r{6.8},  one  can  simplify the
expression for $<\Phi,\Phi>$:
\beb
<\Phi,\Phi> \simeq \int d\alpha (g, \int d\beta d\rho
e^{i\beta_i(\frac{\partial P}{\partial\alpha_i}  \xi  - \frac{\partial
Q}{\partial \alpha_i} \frac{1}{i} \frac{\partial}{\partial\xi})}
e^{i\rho_a (\frac{\partial \Lambda_a}{\partial Q}  \xi  + \frac{\partial
\Lambda_a}{\partial P} \frac{1}{i} \frac{\partial}{\partial\xi})} g) =
\\
\int d\alpha (g,
\prod_{i=1}^k 2\pi
\delta
(\frac{\partial P}{\partial\alpha_i}  \xi  - \frac{\partial
Q}{\partial \alpha_i} \frac{1}{i} \frac{\partial}{\partial\xi})
\prod_{a=1}^M 2\pi \delta(\Xi \Lambda_a) g)
\l{6.10}
\eeb

\subsubsection{Definition of a composed semiclassical state}

We see  that  the  composed  semiclassical  wave  function  \r{6.1} is
specified, if the following properties C1-C2 are satisfied.

C1. {\it  A  manifold   $X(\alpha)   \equiv   (S(\alpha),   Q(\alpha),
P(\alpha))$, $\alpha = ( \alpha_1,...,  \alpha_k)$ is given. It should
obey eq.\r{6.6} (the Maslov "isotropic condition") and belong  to  the
base $\cal X$ of the semiclassical bundle (i.e.  obey eq.\r{6.8}). The
manifold $X(\alpha,\mu) = \lambda_{\mu}X (\alpha)$ should  be  smooth,
its $Q,P$-component $(Q(\alpha,\mu),P(\alpha,\mu))$ should be a smooth
$k+M$-dimensional manifold without self-intersections. }

C2. {\it A function $g\in {\cal S}({\bf R}^{k+n})$ is specified.
}

Note that  the  set  of  parameters  $\alpha$  may  take  values  on a
nontrivial manifold $\Lambda^k$ (such as circle,  torus  etc.)  rather
than ${\bf R}^k$.

The inner  product  of  the  composed semiclassical states is given by
formula \r{6.10}.  All  the  operators   $\frac{\partial   P}{\partial
\alpha_i} \xi   -  \frac{\partial  Q}{\partial  \alpha_i}  \frac{1}{i}
\frac{\partial}{\partial \xi}$, $\Xi\lambda_a$ commute each other.

It is interesting to note that maximal value of dimensionality $k$  is
$n-M$. Namely, there are $M+k$ tangent vectors to the phase space,
\beb
\delta P^{(i)} = \frac{\partial P}{\partial \alpha_i}, \qquad
\delta Q^{(i)} = \frac{\partial Q}{\partial \alpha_i}, \qquad
i= \overline{1,k}.\\
\delta P^{(k+a)} = - \frac{\partial \Lambda_a}{\partial Q_a}, \qquad
\delta Q^{(k+a)} = \frac{\partial \Lambda_a}{\partial P_a}, \qquad
a= \overline{1,M}
\l{6.11a}
\eeb
such that
\beq
\delta P^{(\alpha)} \delta Q^{(\beta)} -
\delta Q^{(\alpha)} \delta P^{(\beta)} = 0.
\l{6.11}
\eeq
The $M+k$-dimensional   plane   $span\{\delta   Q^{(\alpha)},   \delta
P^{(\alpha)}\}$ is called isotropic.  It is known \c{Maslov2,KM}  that
maximal dimensionality of an isotropic plane is $n$.

One can  also  notice  that  there is a gauge freedom in choosing $g$.
Namely, the transformation
\bez
g \to (\Xi \Lambda_a) \chi^a + \Omega[\delta Q^{(i)},  \delta P^{(i)}]
\eta^i
\eez
takes a composed semiclassical state to equivalent.

Let us  formulate  a  definition  of  a  composed  semiclassical state
(cf.\c{Shvedov}). First, introduce auxiliary notions.
Let $L_{k+M}$  be  a  $k+M$-dimensional isotropic plane with a measure
$d\sigma$ being invariant under  shifts.  Let
$(\delta  Q^{(\alpha)},
\delta P^{(\alpha)})$,
$\alpha  =  \overline{1,k+M}$  be  a  basis on
$L_{k+M}$. One can assign then coordinates $\beta_1,...,  \beta_{k+M}$
to any   element   $(\delta   Q,\delta   P)   =  \sum_{\alpha=1}^{k+M}
\beta_{\alpha} (\delta  Q^{(\alpha)},   \delta   P^{(\alpha)})$.   The
measure $d\sigma$  is  presented  as $ad\beta_1 ...  d\beta_{k+M}$ for
some constant $a$. Consider the inner product
\bez
<g_1,g_2>_{L_{k+M}} = a \int d\beta  (g_1,  e^{i\beta_a  \Omega(\delta
Q^{(\alpha)}, \delta   P^{(\alpha)})}   g_2)   =  \int  d\sigma  (g_1,
e^{i\Omega[\delta Q,\delta P]} g_2),
\eez
$g_1,g_2 \in {\cal S}({\bf R}^{k+n})$.  This  definition  is  invariant
under change of basis. We say that
$g \stackrel{L_{k+m}}{\sim} 0$ if $<g,g>_{L_{k+M}} = 0$. Denote
${\cal F}(L_{k+M}) = \overline{{\cal S}({\bf R}^n)/\sim}$.

Let $M^k$ be a $k$-dimensional manifold embedded to $\cal X$,
$\{X(\alpha), \alpha \in M^k\}$,
$d\Sigma$ be a measure on $M^k$. Let property C1 be satisfied. Define
\bez
L_{k+M}(\alpha) =  L_{k+M}(\alpha:M^k) = span \{(\delta Q^{(\alpha)},
\delta P^{(\alpha)}\}.
\eez
for vectors $(\delta Q^{(\alpha)},  \delta P^{(\alpha)})$ of the  form
\r{6.11a}. Plane  $L_{k+M}(\alpha)$  does not depend on the particular
choice of coordinates $\alpha_1,...\alpha_k$.  Introduce the following
measure $d\sigma(\alpha)$ on $L_{k+M}(\alpha)$:
\beq
d\sigma(\alpha) =    \frac{D\Sigma(\alpha)}{D\alpha}    d\beta_1   ...
d\beta_{k+M}.
\l{6.12}
\eeq
Here $(\beta_1,...,\beta_{k+M})$  are coordinates on $L_{k+M}(\alpha)$
which are introduced as
\bez
(\delta Q,  \delta   P)   =   \sum_{\alpha}   \beta_{\alpha}   (\delta
Q^{(\alpha)}, \delta P^{(\alpha)}).
\eez
Definition \r{6.12}   is   invariant   under   change  of  coordinates
\c{Shvedov}.

Introduce the Hilbert bundle $\pi_{M^k}$ as follows.  The base is  the
isotropic manifold $M^k$. The fibre corresponding to the point $\alpha
\in M^k$ is ${\cal H}_{\alpha} = {\cal F}(L_{k+M}(\alpha))$.  Composed
semiclassical states  are  viewed  as sections $Z$ of $\pi_{M^k}$ such
that the inner product
\beq
<(M^k,Z), (M^k,Z)>           =           \int_{M^k}            d\Sigma
<Z(\alpha),Z(\alpha)>_{L_{k+M}(\alpha)}.
\l{6.13}
\eeq
converges.

\subsubsection{Semiclassical transformations of composed semiclassical
states}

Analogously to section 5,  one can apply the operator $e^{-i\hat{O}t}$
to the  composed  semiclassical  state  for  the  quantum   observable
$\hat{O}$ \r{5.i}.  The  wave  function  $\Phi^t = e^{-i\hat{O}t}\Phi$
will satisfy the Cauchy problem \r{5.0}. The substitution
\bez
\Phi^t(q) = c \int d\alpha e^{\frac{i}{h} S^t(\alpha)}
e^{\frac{i}{h} P^t(\alpha) (q\sqrt{h} - Q^t(\alpha))}
g^t(\alpha, q- \frac{Q^t(\alpha)}{\sqrt{h}})
\eez
will give an  asymptotic  solution  of  the  Cauchy  problem  \r{5.0},
provided that   $(S^t,Q^t,P^t)$   satisfies  the  Cauchy  problem  for
eq.\r{5.1}, while eq.\r{5.2} is  valid  for  $g^t(\alpha,\xi)$.  After
differentiation of eq.\r{5.1} with respect to $\alpha$ we find that
\bez
[i\frac{d}{dt} -   \frac{1}{2}  (\Xi^2O)(u_tX);  \Omega(\frac{\partial
O^t}{\partial \alpha_i}, \frac{\partial P^t}{\partial \alpha_i}) = 0.
\eez
Analogously to subsection 5.1,  this implies that  the  inner  product
\r{6.10} (or \r{6.13}) conserves under time evolution.

Equivalence of composed semiclassical states is introduced in the same
way as  equivalence  of   elementary   semiclassical   states.   Since
expression \r{6.1}   is   a   linear   superposition   of   elementary
semiclassical states, all the results concerning equivalence are valid
for the composed semiclassical states as well.

\subsection{Composed semiclassical states in the Dirac approach}

Composed semiclassical  states  can  be also investigated in the Dirac
approach. Consider the wave function \r{6.2}.  Let  us  calculate  the
function $\chi^{\tau}$  being  a  solution  of  the Cauchy problem for
eq.\r{2.i2}. Let us look for it in the following form:
\beq
\chi^{\tau}(\alpha,\xi,\Pi,\overline{\Pi}) =           \varphi^{\tau}
(\alpha,\xi)
\exp[- \overline{\Pi}_a M^a_b(\alpha,\tau) \Pi^b]
\l{6.14}
\eeq
Substituting expression \r{6.14} to eq.\r{2.i2}, one obtains that
\beq
i \dot{\varphi}^{\tau}      =      [\frac{1}{2}      \mu_a      (\Xi^2
\Lambda_a)(Q^{\tau},P^{\tau}) +                                  \mu_a
\Lambda_a^{(1)}(Q^{\tau},P^{\tau})] \varphi^{\tau},
\l{6.15}
\eeq
while
\beq
\dot{M}^a_b = - \delta^a_b - \mu_c U^a_{dc}(Q^{\tau},P^{\tau}) M^d_b.
\l{6.16}
\eeq
Integrating function \r{6.14} over Grassmannian variables, we find
\bez
g^{\tau}(\alpha,\xi) = det M \varphi^{\tau}(\alpha,\xi).
\eez
We are interested in $g^{\tau}$ at $\tau = -1$.

One can    notice    that   for   $X(\alpha,\mu)   =   (S(\alpha,\mu),
Q(\alpha,\mu), P(\alpha,\mu))$ one has
\bez
X(\alpha,\mu) = \lambda_{-\mu} X(\alpha),
\eez
while
\bez
\varphi^{-1} (\alpha,\xi)  =   V^0_{-\mu}(\lambda_{-\mu}   X\gets   X)
g(\alpha).
\eez
Let us  investigate  the  matrix $M$.  It follows from eq.\r{3.8} that
eq.\r{6.16} can be rewritten as
\bez
\frac{dM}{d\tau} + (Ad_X(-\mu \tau))^{-1}  \frac{d}{d\tau}  (Ad_X(-\mu
\tau)) = -1,
\eez
so that
\bez
Ad_X(-\mu\tau) M(\tau) = - \int_0^{\tau} d\tau' Ad_X (-\mu\tau').
\eez
For the case $\tau = -1$, one takes this relation to the form
\beq
M(-\mu,X) = (Ad_X \mu)^{-1} \int_0^1 d\tau Ad_X(\mu\tau).
\l{6.16a}
\eeq
Eq.\r{6.16a} can be simplified. First, notice
\bez
(Ad_X \mu)^{-1} Ad_X (\mu(1-\tau)) = Ad_{\lambda_{-\mu}X} (-\mu\tau),
\eez
so that
\bez
M(-\mu,X) = \int_0^1 d\tau Ad_{\lambda_{-\mu}X} (-\mu\tau).
\eez
It follows from eq.\r{3.24} that
\beq
M^a_b = \frac{\delta \sigma^a}{\delta \mu^b}.
\l{6.16b}
\eeq
Here $\delta\sigma$ and $\delta \mu$ are related as follows,
\bez
\lambda_{-\delta\sigma} \lambda_{-\mu}  X =
\lambda_{-\mu-\delta \rho} X.
\eez
Thus,
\bez
g(\alpha,\mu) =  det  M(-\mu,  X(\alpha))  V^0_{-\mu}  (\lambda_{-\mu}
X(\alpha) \gets X(\alpha)) g(\alpha).
\eez
The measure  $det  M  d\mu$  resembles  the  invariant  measure on the
quasigroup constructed in \c{Batalin}.

Analogously to expression \r{6.1},  the wave function \r{6.2} will not
vary in the leading order in $h$ under transformation
\beq
g \Rightarrow g + \Omega(\frac{\partial Q}{\partial\alpha_i},
\frac{\partial P}{\partial \alpha_i}) \chi^i  +  \Omega(\frac{\partial
Q}{\partial \mu_a}, \frac{\partial P}{\partial\mu_a}) \zeta^a.
\l{6.17}
\eeq
Let us show that the Dirac wave  function  is  invariant  under  gauge
transformations of  $\Phi$  and  satisfies  the  constraint
conditions
\beq
\hat{\Lambda}_a^+ \Psi = 0.
\l{6.18}
\eeq
It is  sufficient to check these properties for the case of elementary
semiclassical states.

\subsubsection{Invariance of   Dirac   wave   function   under   gauge
transformations}

Let
\bez
\Phi(q) =  c(K_X^h  g)(q)  \equiv  c  e^{\frac{i}{h}S}  e^{\frac{i}{h}
P(q\sqrt{h} -Q)} f(q- \frac{Q}{\sqrt{h}})
\eez
be a wave  packet  wave  function  for  $X=(S,Q,P)$.  Then  the  Dirac
function will be of the form
\beq
\Psi =  c  \int  d\mu K^h_{\lambda_{-\mu}X} V^0_{-\mu}(\lambda_{-\mu}X
\gets X) g \det M (-\mu,X).
\l{6.19}
\eeq
Consider the gauge transformation
\bez
X \to  \lambda_{-\mu}X;  \qquad g \to V^0_{-\nu}(\lambda_{-\nu}X \gets
X) g.
\eez
For the transformed semiclassical state,  the Dirac wave function will
take the form
\beq
\tilde{\Psi} =   c\int   d\mu   K^h_{\lambda_{-\mu}   \lambda_{-\nu}X}
V^0_{-\mu} (\lambda_{-\mu}  \lambda_{-\nu}X   \gets   \lambda_{-\nu}X)
V^0_{-\nu} (\lambda_{-\nu}X \gets X)g \det M(-\mu, \lambda_{-\nu}X)
\l{6.20}
\eeq
Making use of the gauge function \r{6.17} (if  necessary),  one  takes
formula \r{6.20} to the form
\beq
\tilde{\Psi} \simeq  c  \int  d\mu K^h_{\lambda_{-\rho} X} V^0_{-\rho}
(\lambda_{-\rho}X \gets X) g \det M(-\mu,\lambda_{-\nu}X)).
\l{6.21}
\eeq
Here
\beq
\lambda_{-\rho} X = \lambda_{-\mu} \lambda_{-\nu} X
\l{6.22}
\eeq
for $\rho =  \rho(\mu,\nu,  X)$.  Expressions  \r{6.19}  and  \r{6.21}
coincide, provided that
\beq
\frac{D\mu}{D\rho} \det M(-\mu, \lambda_{-\nu}X) = \det M(-\rho,X).
\l{6.23}
\eeq
Relation \r{6.23} is a corollary of \r{6.16b}. Namely,
\bey
\frac{D\mu}{D\rho} = \det \frac{\delta \mu}{\delta \rho};
\qquad
\lambda_{-\mu-\delta \mu}         \lambda_{-\nu}          X          =
\lambda_{-\rho-\delta\rho} X;
\\
\det M(-\mu, \lambda_{-\nu}X) = \det \frac{\delta \nu}{\delta \mu};
\qquad
\lambda_{-\delta \nu} \lambda_{-\mu} \lambda_{-\nu} X =
\lambda_{-\mu-\delta \mu} \lambda_{-\nu} X.
\\
\det M(-\rho, X) = \det \frac{\delta \nu}{\delta \rho};
\qquad
\lambda_{-\delta \nu}  \lambda_{-\rho} X =
\lambda_{-\rho-\delta \rho}  X.
\eey
Thus, eq.\r{6.23}  is  satisfied,  and  the  Dirac  wave  function  is
invariant under gauge transformation.

\subsubsection{Constraint conditions}

To check relation \r{6.18}, it is more convenient to justify that
\beq
\exp [i\hat{\Lambda}_a^+ \nu_a] \Psi = \Psi.
\l{6.24}
\eeq
The left-hand side of eq.\r{6.24}
\bey
\tilde{\Psi} = c\int d\mu K_{\lambda_{-\nu} \lambda_{-\mu} X} V^0_{-\nu}
(\lambda_{-\nu} \lambda_{-\mu} X \gets \lambda_{-\mu} X)
V^0_{-\mu} (\lambda_{-\mu} X \gets X) g
\det M(-\mu, X)
\\ \times
\exp[-i\int_0^{-1} d\tau \nu_a \{
\Lambda_a^{(1)*} (\lambda_{\nu\tau} \lambda_{-\mu} X) -
\Lambda_a^{(1)} (\lambda_{\nu\tau} \lambda_{-\mu} X) \}]
\eey
can be presented as
\beq
\tilde{\Psi} = c\int d\rho \frac{D\mu}{D\rho}
K_{\lambda_{-\rho}X} V^0_{-\rho} (\lambda_{-\rho}X \gets X) g
\det (-M(-\mu,X))
\exp[-\int_0^1     d\tau     \nu_a     U^d_{da}
(\lambda_{\nu\tau} \lambda_{-\mu} X)]
\l{6.25}
\eeq
Here
\beq
\lambda_{-\nu} \lambda_{-\mu} X = \lambda_{-\rho} X
\l{6.26}
\eeq
and property \r{2.5c} is  taken  into  account.  Moreover,  eq.\r{3.8}
implies that
\bez
\exp[-\int_0^1     d\tau     \nu_a     U^d_{da}
(\lambda_{\nu\tau} \lambda_{-\mu} X)] =
\exp [- \log \det Ad_{-\lambda_{-\mu}X} (-\nu\tau)|_0^{-1}]
= (\det Ad_{-\lambda_{-\mu}X} (-\nu))^{-1}.
\eez
Eqs.\r{6.25} and \r{6.19} coincide if
\beq
\frac{D\mu}{D\rho} \det M(-\mu,X)
(\det Ad_{-\lambda_{-\mu}X} (-\nu))^{-1} =
\det M(-\rho, X).
\l{6.27}
\eeq
Here $\mu(\rho,\nu)$ is defined from eq.\r{6.26}.

Moreover, eq.\r{6.27} is satisfied, since
\bey
\frac{D\mu}{D\rho} = \det \frac{\delta \mu}{\delta \rho};
\qquad
\lambda_{-\nu} \lambda_{-\mu-\delta \mu} X =
\lambda_{-\rho-\delta\rho} X;
\\
\det M(-\mu, X) = \det \frac{\delta \delta \sigma}{\delta \mu};
\qquad
\lambda_{-\delta\sigma} \lambda_{-\mu} X =
\lambda_{-\mu - \delta \mu} X.
\\
(\det Ad_{\lambda_{-\mu} X}\nu)^{-1} =
\det \frac{\delta \delta \nu}{\delta \sigma};
\qquad
\lambda_{-\nu} \lambda_{-\delta\sigma} \lambda_{-\mu} X =
\lambda_{-\delta \nu} \lambda_{-\nu} \lambda_{-\mu} X;
\\
\det M(-\rho, X) = \det \frac{\delta \nu}{\delta \rho};
\qquad
\lambda_{-\delta \nu}  \lambda_{-\rho} X =
\lambda_{-\rho-\delta \nu}  X.
\eey
Constraint \r{6.18} is checked.

\section{Discussion}

Let us discuss the obtained results.

We have started from  the  quantum  theory  of  the  system  with  $M$
first-class constraints  depending  on  the  small  parameter  of  the
semiclassical expansion  $h$  according  to  eq.\r{2.2x}.  Notions  of
elementary and composed semiclassical states have been introduced.

There are   different  ways  to  quantize  a  constrained  system.  To
investigate elementary  semiclassical  states,  the  most   convenient
quantization technique is the refined algebraic quantization approach.
Elementary semiclassical states are specified by sets  $(X,f)$,  where
$X=(S,P,Q)$ is  a  classical state belonging to the constraint surface
$\Lambda_a(Q,P)=0$, while $f$ is  a  quantum  state  in  the  external
background $X$. The quantum wave function depends on $h$ as \r{2.12}.

It has  been  shown  that  the  condition  $\Lambda_a(Q,P)=0$  is very
important. If  it  is  not   satisfied,   the   norm   of   elementary
semiclassical state is exponentially small.

The inner  product  of semiclassical states has been calculated in the
semiclassical approximation (eq.\r{2.23}).  However,  this formula  is
valid only  if  the linearized constraints $\Xi\Lambda_a$ (eq.\r{A.5})
are independent.   The   case   of   linearly   dependent    operators
$\Xi\Lambda_a$ can be investigated as follows.  One should choose such
a basis in the Lie algebra of constraints  that  $\Xi\Lambda_A  =  0$,
$A=\overline{1,D}$, $0  <  D  \le  M$,  while $\Xi\Lambda_{D+\alpha}$,
$1 \le \alpha \le M-D$ are linearly independent.  Then one writes down
formula \r{2.4m},       rescale       $\mu^{D+\alpha}      \Rightarrow
\sigma^{\alpha}\sqrt{h}$, $\mu^A \Rightarrow \rho^A$.  Analogously  to
subsection 2.2, one finds
\bez
<\Phi,\Phi> \simeq h^{D/2} |c|^2 \int d\rho d\sigma J(\rho,0)
(f, e^{-i\rho^A \frac{1}{2} \Xi^2  \Lambda_A  -  i\sigma^{\alpha}  \Xi
\Lambda_{D+\alpha}} f).
\eez
We see that if some of linearized constraints vanish,  one should take
into account the quadratic part of them.

Since the inner product \r{2.23} has appeared to  be  degenerate,  one
should say  that two semiclassical wave functions corresponding to the
same $X$ are  equivalent  if  their  difference  has  zero  norm  (for
example, of  the form $(\Xi \Lambda_a)(X) \chi^a$).  Thus,  it is more
correct to say that elementary semiclassical states are specified by a
set of a classical state $X$ and a class of equivalence $[f]$.  Set of
all elementary semiclassical states forms a semiclassical bundle  with
the base $\{ (S,P,Q) | \Lambda_a(Q,P) = 0\}$  and fibres being spaces
of states $[f]$.  Elementary semiclassical states are then  points  on
the semiclassical bundle.

An important  property  of  theories  of  constrained systems is gauge
invariance. In the refined algebraic quantization approach, this means
that quantum  states $\Phi$ and $e^{-i\tau \mu^a\hat{\Lambda}_a} \Phi$
are equivalent.  For  the  elementary   semiclassical   state   $\Phi$
specified by $(X,[f])$, the wave function
$e^{-i\tau \mu^a\hat{\Lambda}_a} \Phi$ calculated  explicitly  in  the
semiclassical approximation   has   appeared   to   be  an  elementary
semiclassical state $(\lambda_{\mu\tau} X, V(\lambda_{\mu\tau} X \gets
X) f)$.  We see that gauge group acts on the semiclassical bundle,  so
that some of elementary semiclassical states are gauge-equivalent. The
group and  quasigroup properties of semiclassical transformations have
been investigated.

Elementary semiclassical states can be also  investigated  within  the
Dirac approach  discussed in section 6.  The wave function \r{6.19} is
specified then by an  $M$-dimensional  surface  on  the  semiclassical
bundle. The surface can be interpretted as a gauge orbit.

Composed semiclassical states are introduced as superpositions \r{6.1}
of elementary  semiclassical  states.   In   the   refined   algebraic
quantization approach,  they are specified by $k$-dimansional surfaces
on the  semiclassical  bundle  $(X(\alpha),g(\alpha))$  with  $\alpha$
being a $k$-dimensional variable, $X(\alpha)$ be an $\alpha$-dependent
classical state, $g(\alpha)$ be an $\alpha$-dependent quantum state in
the external  background  $X(\alpha)$.  The  inner product of composed
semiclassical states has been evaluated (eq.\r{6.10}).  In  the  Dirac
approach, the   dimensionality   of   the   surface  embedded  to  the
semiclassical bundle is $M+k$.

Evolution transformations of  elementary  and  composed  semiclassical
states have  been investigated,  provided that the quantum Hamiltonian
depends on the small parameter as \r{5.i}. Semiclassical state $(X,f)$
is taken to $(u_tX,  U_t(u_tX \gets X)f)$. It has been also shown that
gauge-equivalent semiclassical states are taken to gauge equivalent.

The obtained  results  can  be  used  for  finding  the  semiclassical
spectrum of  a  semiclassical  observable.  One should consider such a
semiclassical initial condition  for  the  Schrodinger  equition  that
conserves its  form  under  time  evolution.  This means that manifold
$(P^t(\alpha),Q^t(\alpha))$ should     be     gauge-equivalent      to
$(P(\alpha),Q(\alpha))$. Certain  conditions  on  $f^t$  can  be  also
obtained. For  example,  one  can  choose  a  stationary  point  of  a
Hamiltonian system  or a periodic trajectory $(P(t),Q(t))$ as an 0- or
1-dimensional isotropic manifold  and  obtain  a  static  or  periodic
soliton quantization theory \c{soliton}.

We have  noticed  that  $k$-dimensional  isotropic  manifolds  in  the
refined algebraic quantization theory correspond to $k+M$-dimensional
isotropic manifolds   in   the   Dirac   approach.  It  is  well-known
\c{Maslov1} that maximal dimensionality of an  isotropic  manifold  is
$n$; this  corresponds  to  the semiclassical WKB theory.  We see that
WKB-method can be developed for the Dirac quantization approach  only,
while the  wave  packet method can be applied to the refined algebraic
quantization only.  Main formulas of  the  WKB  theory  in  the  Dirac
approach may   be   obtained  without  using  integral  representation
\r{6.19}. Namely,  the WKB wave function has  the  following  explicit
form according   to  subsubsection  6.1.1.  If  $k+M=n$,  in  "general
position" case for all $q$ there exists such $\overline{\alpha}$  that
$q=h^{-1/2} Q(\overline{\alpha})$. Eq.\r{6.1xx} reads:
\bez
\Psi(q) = \varphi(q\sqrt{h}) \exp(\frac{i}{h} S(q\sqrt{h}))
\eez
for some  $\varphi$  and  $S$.  Eq.\r{6.6}  implies that the isotropic
manifold $(P(\alpha),Q(\alpha))$ coincides with
\bez
\{ (\frac{\partial S}{\partial X_i}, X_i) | X \in {{\bf R}}^n \}
\eez
The Dirac condition \r{6.18} can be rewritten as
\bez
\Lambda_a^*(X, \frac{\partial      S}{\partial      X}      -       ih
\frac{\partial}{\partial X}) \varphi(X) = 0.
\eez
In the leading order in $h$, one finds that the isotropic
manifold lies on the constraint  surface,  the  next-to-leading  order
gives us first-oder equations on $\varphi$.

However, the  main  difficulty  of the WKB-approach is that the Cauchy
problem Hamiltonian-Jacobi equation may have no solutions (this  means
that the  projection  of  the isotropic manifold to the plane $P=0$ is
not unique).  For  such  cases,  the  more  complicated  semiclassical
methods such  as  the  Maslov  canonical operator approach \c{Maslov1}
should be used.  One of possible formulations of this method is  based
on the integral representation \r{6.1}.

The discussed  derivation  of the semiclassical theory is based on the
quantum theory:  it was used as a starting point,  then the notions of
semiclassical states,   observables   and  evolution  was  introduced.
However, semiclassical mechanics can be also viewed as a first step of
quantization of classical theory.  Indeed,  it has been shown that all
objects of   the   semiclassical   theory   (inner   product,    gauge
transformation, semiclassical        observables,        semiclassical
transformation) are expressed in terms of classical variables.

\section*{Acknowledgments}

The author  is  indebted  to  J.Klauder,  D.Marolf  and V.P.Maslov for
helpful discussions.
This work  was supported by the Russian Foundation for Basic Research,
projects 99-01-01198 and 01-01-06251.

\appendix

\section{Some properties of Weyl quantization}

Let us review some results of the theory of  Weyl  quantization  (see,
for example, \c{B,KM}).

There are different ways to define a notion of a function of operators
$q$ and $-i\partial/\partial q$.  One can put the coordinates  to  the
right and momenta to the left and wise versa. An alternative way is to
use the Weyl ordering.  Let $f(q,p)$ be an arbitrary function of $q\in
{\bf R}^n$,  $p  \in  {\bf R}^n$.  Consider its expansion as a Fourier
integral,
\beq
f(q,p) = \int d\alpha d\beta \tilde{f}(\alpha,\beta)  e^{i\alpha  q  +
i\beta p}.
\l{A.0}
\eeq
By definition, set
\beq
W[f] \equiv f(q, -i \frac{\partial}{\partial q}) =
\int d\alpha d\beta \tilde{f}(\alpha,\beta)
e^{i\alpha  q  +
i\beta (-i \frac{\partial}{\partial q})}.
\l{A.1}
\eeq
The operator $W[f]$ is called a Weyl quantization of the function $f$,
while $f$ is called a Weyl symbol of the operator $W[f]$. The exponent
$e^{i\alpha  q  +
i\beta (-i  \frac{\partial}{\partial  q})}$ entering to eq.\r{A.1} can
be defined with the help of the Baker-Hausdorff formula,
\bez
(e^{i\alpha\hat{q} + i\beta\hat{p}} \psi)(x) =
(e^{i\alpha\hat{q}} e^{i\beta\hat{p}} e^{\frac{i}{2} \alpha \beta}
\psi)(x) =
e^{i\alpha (x+ \frac{\beta}{2})}
(e^{\beta \frac{\partial}{\partial x}}\psi) (x) =
e^{i\alpha (x+ \frac{\beta}{2})}
\psi(x+\beta).
\eez

{\bf Proposition  A.1.} {\it The Weyl symbol of the product $W[f]W[g]$
is
\beq
(f*g)(q,p) =
\int \frac{dk_1 dk_2 dx_1 dx_2}{(2\pi)^{2n}} f(q+x_1,  p+
\frac{k_2}{2}) g(q+x_2, p- \frac{k_1}{2}) e^{-ik_1x_1 - ik_2x_2}
\l{A.2}
\eeq
}

{\bf Proof.} One has
\bez
W[f] W[g] = \int
d\alpha d\beta d\alpha' d\beta'
\tilde{f}(\alpha,\beta)
\tilde{g}(\alpha',\beta')
e^{i\alpha  \hat{q}  + i\beta \hat{p}}
e^{i\alpha'  \hat{q}  + i\beta' \hat{p}}.
\eez

Making use of the Baker-Hausdorff formula,  we take this  relation  to
the form
\bez
W[f]W[g] =
d\alpha'' d\beta'' d\alpha' d\beta'
\tilde{f}(\alpha'',\beta'')
\tilde{g}(\alpha',\beta')
e^{i(\alpha'' + \alpha')  \hat{q}  + i(\beta'' + \beta') \hat{p}}
e^{\frac{i}{2} (\alpha'' \beta' - \alpha' \beta'')}
\eez
here the redefining $\alpha \to\alpha'$, $\beta\to\beta'$ is made.
Therefore, the Fourier transformation of $f*g$ is
\bez
(\tilde{f*g})(\alpha,\beta) = \int d\alpha' d\beta' \tilde{f}(\alpha -
\alpha', \beta-\beta') \tilde{g}(\alpha',\beta')
e^{\frac{i}{2} ((\alpha-\alpha') \beta' - \alpha' (\beta - \beta'))}
\eez
Making use  of  formula  \r{A.0} and analogous formula for the inverse
Fourier transformation,  we obtain relation  \r{A.2}.  Proposition  is
proved.

Let $f$ and $g$ depend on the small parameter $h$ as follows,
\bez
f = F(q\sqrt{h},p\sqrt{h}), \qquad g = G(q\sqrt{h},p\sqrt{h}).
\eez
then
\bez
F(\hat{q}\sqrt{h},\hat{p}\sqrt{h}) G(\hat{q}\sqrt{h}, \hat{p}\sqrt{h})
= H_h (\hat{q}\sqrt{h}, \hat{p}\sqrt{h})
\eez
with
\bez
H_h(Q,P) =   \int   \frac{dk_1   dk_2   dx_1  dx_2}{(2\pi)^{2n}}
F(Q- \frac{\sqrt{h}}{2} x_1,  P +
\frac{\sqrt{h} k_2}{2})
G(Q+\sqrt{h} x_2, P + \sqrt{h} {k_1}) e^{-ik_1x_1 - ik_2x_2}
\eez
Here the rescaling $x_1 \to -x_1/2$, $k_1 \to -2k_1$ is performed. Let
us simplify this expression. One has
\bez
F(Q- \frac{\sqrt{h}}{2} x_1,  P + \frac{\sqrt{h} k_2}{2})
e^{-ik_1x_1 - ik_2x_2} =
F(Q- \frac{i\sqrt{h}}{2} \frac{\partial}{\partial k_1},
P + \frac{i\sqrt{h}}{2} \frac{\partial}{\partial x_2})
e^{-ik_1x_1 - ik_2x_2}
\eez
Integrating exponent over $k_2$ and $x_1$, we obtain $\delta$-function
$\delta(x_2) \delta(k_1)$. Integration by parts gives us the following
expression
\bez
H_h(Q,P) = F(Q + \frac{i\sqrt{h}}{2} \frac{\partial}{\partial k_1};
P -      \frac{i\sqrt{h}}{2}      \frac{\partial}{\partial       x_2})
G(Q+\sqrt{h}x_2; P+\sqrt{h}k_1)|_{x_2=0, k_1=0}.
\eez
We obtain the following proposition.

{\bf Proposition A.2.} {\it The following relation takes place:
\bey
H_h = FG + \frac{ih}{2} \left(
\frac{\partial F}{\partial Q} \frac{\partial G}{\partial P}
- \frac{\partial F}{\partial P} \frac{\partial G}{\partial Q}
\right) +\\
\frac{h^2}{4} \left(
- \frac{1}{2} \frac{\partial^2F}{\partial Q_i\partial Q_j}
\frac{\partial^2G}{\partial P_i\partial P_j}
- \frac{1}{2} \frac{\partial^2G}{\partial Q_i\partial Q_j}
\frac{\partial^2F}{\partial P_i\partial P_j}
+ \frac{\partial^2F}{\partial Q_i\partial P_j}
\frac{\partial^2G}{\partial P_i\partial Q_j}
\right)
+ o(h^2).
\eey
}

Consider now  the  operator  $F(Q+\xi\sqrt{h},  P + \frac{\sqrt{h}}{i}
\frac{\partial}{\partial\xi}$ which can be defined as
\bez
F(Q+\xi\sqrt{h}, P + \frac{\sqrt{h}}{i} \frac{\partial}{\partial \xi})
= \int     d\alpha     d\beta    \tilde{F}(\alpha,\beta)    e^{i\alpha
(Q+\xi\sqrt{h}) +      i\beta      (P       +       \frac{\sqrt{h}}{i}
\frac{\partial}{\partial\xi})}.
\eez
We obtain the following proposition.

{\bf Proposition A.3.} {\it The following relation is satisfied:
\beq
F(Q+\xi\sqrt{h}, P + \frac{\sqrt{h}}{i} \frac{\partial}{\partial \xi})
= e^{\sqrt{h} (\xi \frac{\partial}{\partial Q}
+ \frac{1}{i} \frac{\partial}{\partial \xi} \frac{\partial}{\partial P}
)} F(Q,P).
\l{A.4}
\eeq
}

Let us expand the operator expression \r{A.4} into a series. One has
\bez
F(Q+\xi\sqrt{h}, P + \frac{\sqrt{h}}{i} \frac{\partial}{\partial \xi})
= F(Q,P) + \sqrt{h} (\Xi F)(Q,P) + \frac{h}{2} (\Xi^2 F)(Q,P) + ...
\eez
with
\bez
\Xi =    \xi     \frac{\partial}{\partial     Q}     +     \frac{1}{i}
\frac{\partial}{\partial \xi} \frac{\partial}{\partial P}.
\eez
Explicitly, one has
\beb
\Xi F   =   \xi_i   \frac{\partial   F}{\partial  Q_i}  +  \frac{1}{i}
\frac{\partial}{\partial \xi_i} \frac{\partial F}{\partial P_i},\\
\frac{1}{2} (\Xi^2  F) =
\frac{1}{2} \xi_i \frac{\partial^2F}{\partial Q_i \partial Q_j} \xi_j +
\frac{1}{2} \xi_i \frac{\partial^2F}{\partial Q_i \partial P_j}
(-i \frac{\partial}{\partial \xi_j}) +
\frac{1}{2} (-i \frac{\partial}{\partial \xi_i})
\frac{\partial^2F}{\partial P_i \partial Q_j} \xi_j
+
\frac{1}{2} (-i \frac{\partial}{\partial \xi_i})
\frac{\partial^2F}{\partial P_i \partial Q_j}
(-i \frac{\partial}{\partial \xi_j}).
\l{A.5}
\eeb

Let us investigate some properties of the operators $\Xi F$ and $\Xi^2
F$.

{\bf Proposition A.4.} {\it  The following properties are obeyed:}
\bez
\Xi(AB) = \Xi A \cdot B + A \cdot \Xi B;
\eez
\bez
\frac{1}{2}\Xi^2(AB) =
\frac{1}{2} \Xi^2A \cdot B + \frac{1}{2} \Xi A \cdot \Xi B
+ \frac{1}{2} \Xi B \cdot \Xi A + \frac{1}{2} A \cdot \Xi^2 B;
\eez
\bez
[\Xi A; \Xi B] = -i \{A;B\} = -i
(\frac{\partial A}{\partial P} \frac{\partial B}{\partial Q}
- \frac{\partial A}{\partial Q} \frac{\partial B}{\partial P}).
\eez

The proof is by direct calculations.

For each function $A(Q,P)$ introduce the Hamiltonian vector field
\beq
\delta_A =
\frac{\partial  A}{\partial  P_i}  \frac{\partial}{\partial Q_i }
- \frac{\partial  A}{\partial  Q_i}  \frac{\partial}{\partial P_i }
\l{A.6}
\eeq
Obviously,
\bez
[\delta_A; \delta_B] = \delta_{\{A,B\}}.
\eez
Let us investigate properties of the operators
\beq
i\delta_A - \frac{1}{2} \Xi^2A.
\l{A.7}
\eeq

{\bf Proposition   A.5.}   {\it  The  operators  \r{A.7}  satisfy  the
following properties,}
\bez
[ i\delta_A - \frac{1}{2} \Xi^2A; \Xi B] = i\Xi\{A;B\};
\eez
\bez
[ i\delta_A - \frac{1}{2} \Xi^2A;
i\delta_B - \frac{1}{2} \Xi^2B] = i
( i\delta_{\{A;B\}} - \frac{1}{2} \Xi^2\{A;B\}).
\eez

The proof is by direct usage of formulas \r{A.5}, \r{A.6}.

\section{Types of semiclassical wave functions}

The most popular semiclassical approach to quantum  mechanics  is  the
WKB-approach. It is the following. One considers the initial condition
for the equation
\beq
ih \frac{\partial\Psi_t      (X)}{\partial      t}=      h(X,      -ih
\frac{\partial}{\partial X}) \Psi_t(X), \qquad
X \in {\bf R}^n,
\l{B.0}
\eeq
which depends on the small parameter $h$ as follows,
\beq
\Psi_0(X) = \varphi_0(X) e^{\frac{i}{h} S_0(X)},
\l{B.1}
\eeq
where $S_0(X)$ is a real function.  The WKB-result \c{Maslov1} is that
the solution  of  the  Cauchy  problem at time moment $t$ has also the
form \r{B.1} up to $O(h)$,
\bez
\Psi_t(X) = \varphi_t(X) e^{\frac{i}{h} S_t(X)} + O(h).
\eez
Equations for $\varphi_t$ and $S_t$ can be obtained.

However, we are not  obliged  to  choose  the  initial  condition  for
eq.\r{B.0} in  a  form  \r{B.1}.  There  are  other  substitutions  to
eq.\r{B.1} that conserve their form under time evolution as $h\to  0$.
For example,   consider   the   Maslov   complex-WKB   wave   function
\c{Maslov2},
\beq
\Psi_0(X) =  const   e^{\frac{i}{h}S_0}   e^{\frac{i}{h}   P_0(X-Q_0)}
f_0(\frac{X-Q_0}{\sqrt{h}})
\l{B.2}
\eeq
corresponding to the wave packet with uncertainties of  the  coordinate
and momentum of the order $O(\sqrt{h})$. Formula \r{B.2} specifies the
classical particle  with  classical  coordinate  $Q_0$  and  classical
momentum $P_0$.  The  function  $f_0$  specifies the shape of the wave
packet: we see that semiclassical  mechanics  is  indeed  richer  than
classical since there are no analogs of $f_0$ in classical mechanics.

It happens that the initial condition \r{B.2} conserves its form under
time evolution \c{Maslov2},
\beq
\Psi_t(X) \simeq  const   e^{\frac{i}{h}S_t}   e^{\frac{i}{h}   P_t(X-Q_t)}
f_t(\frac{X-Q_t}{\sqrt{h}})
\l{B.3}
\eeq
up to $O(\sqrt{h})$.  The phase factor $S_t$ is the action  along  the
classical trajectory, $P_t,Q_t$ obey the classical Hamiltonian system.
For the function $f_t$ specifying time evolution of the  form  of  the
wave packet,  the Schrodinger equation with a time-dependent quadratic
Hamiltonian is obtained \c{Maslov2,MS}.

The wave function \r{B.1}  rapidly  oscillates  with  respect  to  all
variables $X \in {\bf R}^n$.  The wave packet \r{B.2} rapidly damps at
$X - Q_0 >> O(\sqrt{h})$. One should come to the conclusion that there
exists a  wave  function  asymptotically  satisfying  eq.\r{B.0} which
oscillates with respect to one  group  of  variables  and  damps  with
respect to  other  variables.  Construction of such states is given in
the Maslov  theory  of  Lagrangian   manifolds   with   complex   germ
\c{Maslov2}. Let $\alpha \in {\bf R}^k$,  $(P(\alpha),  Q(\alpha)) \in
{\bf R}^{2n}$ be a $k$-dimensional  surface  in  the  $2n$-dimensional
phase space, $S(\alpha)$ be a real function, $f(\alpha,\xi)$, $\xi \in
{\bf R}^n$ be a smooth function. Set $\Psi(X)$ to be not exponentially
small if  and  only  if  the  distance  between  point $X$ and surface
$Q(\alpha)$ is of the order $\le O(\sqrt{h})$. Otherwise, set $\Psi(X)
\simeq 0$.  If $\min_{\alpha} |X-Q(\alpha)| = |X-Q(\overline{\alpha})|
= O(\sqrt{h})$, set
\beq
\Psi(X) = const e^{\frac{i}{h}S (\overline{\alpha})}
e^{\frac{i}{h}   P (\overline{\alpha})(X-Q(\overline{\alpha}) )}
f(\overline{\alpha}, \frac{X-Q(\overline{\alpha})}{\sqrt{h}})
\l{B.8}
\eeq
One can note that wave functions \r{B.1} and \r{B.2} are partial cases
of the  wave  function  \r{B.8}.  Namely,  for  $k=0$   the   manifold
$(P(\alpha), Q(\alpha))$  is  a  point,  so  that the function \r{B.8}
coincides with  \r{B.2}.  Let  $k=n$.  If  the  surface   $(P(\alpha),
Q(\alpha))$ is in the general position, for $X$ in some domain one has
$X=Q(\overline{\alpha})$ for some $\overline{\alpha}$,  so that  $\Psi
(X) =         const        e^{\frac{i}{h}        S(\overline{\alpha})}
f(\overline{\alpha},0)$ is a WKB-function.

The lack   of   formula   \r{B.8}   is   that   the   dependence    of
$\overline{\alpha}$ on  $X$ is implicit and too complicated.  However,
under certain   conditions   formula   \r{B.8}   is    invariant    if
$\overline{\alpha}$ is   shifted   by   a   quantity   of   the  order
$O(\sqrt{h})$. Such conditions are
\beq
\frac{\partial S}{\partial \overline{\alpha}_i}
= P \frac{\partial Q}{\partial \overline{\alpha}_i}; \\
\l{B.9}
\eeq
\beq
(\xi \frac{\partial P}{\partial \overline{\alpha}_i}
- \frac{1}{i} \frac{\partial}{\partial \xi}
\frac{\partial Q}{\partial \overline{\alpha}_i} f = 0.
\l{B.10}
\eeq
The form  \r{B.8}  of  the  semiclassical state appeared in the Maslov
theory of Lagrangian manifolds with complex germ is not convenient for
quantum field  theory.  It  is  much  more  suitable  to  consider the
superposition
\beq
\Psi(X) = const \int d\alpha e^{\frac{i}{h}S ({\alpha})}
e^{\frac{i}{h}   P ({\alpha})(X-Q({\alpha}) )}
f({\alpha}, \frac{X-Q({\alpha})}{\sqrt{h}})
\l{B.11}
\eeq
of states   \r{B.1}.  Partial  cases  of  such  a  superposition  were
considered in \c{super};  general case is investigated in  \c{MS,MS4}.
The case  of  semiclassical mechanics in abstract spaces is considered
in \c{Sh2}.

It happens that the integral  \r{B.11}  approximately  coincides  with
\r{B.8}, provided that condition \r{B.9} is satisfied and
\bez
f(\overline{\alpha},\xi) = \prod_{s=1}^k (2\pi \delta(
\frac{\partial P}{\partial \overline{\alpha}_s} \xi
- \frac{\partial Q}{\partial \overline{\alpha}_s}
\frac{1}{i} \frac{\partial}{\partial \xi}
) )
g(\overline{\alpha},\xi)
\eez
Thus, condition \r{B.10} is automatically satisfied.

We see  that  the  wave-packet  wave function \r{B.3} may be viewed as
"elementary" semiclassical states,  while wave functions  appeared  in
the theory  of  Lagrangian  manifolds with complex germ (including WKB
functions) can  be  considered  as  superpositions   of   "elementary"
semiclassical states.

\end{document}